\documentclass[3p, twocolumn]{elsarticle}
\biboptions{sort&compress}
\usepackage[ruled,vlined]{algorithm2e}
\usepackage[breaklinks=true]{hyperref}
\usepackage[anythingbreaks]{breakurl}
\usepackage{amsthm}
\usepackage{amsmath}
\usepackage{amsfonts}
\usepackage{amssymb}
\usepackage{amsfonts}
\usepackage{amsthm}
\usepackage[capitalize]{cleveref}
\newtheorem{theorem}{Theorem}

\newtheorem{example}{Example}[section]

\Crefname{conjecture}{Conjecture}{Conjectures} % capitalized version

\Crefname{proposition}{Proposition}{Propositions} % capitalized version
\newtheorem{definition}{Definition}[section]
\Crefname{definition}{Definition}{Definitions} % capitalized version

\Crefname{lemma}{Lemma}{Lemmas} % capitalized version

\Crefname{remark}{Remark}{Remarks} % capitalized version
\usepackage{dirtytalk}
\usepackage{blkarray}
\usepackage{enumitem}
\usepackage{epsfig}
\usepackage{bbm}
\usepackage{todonotes}
\usepackage{tabularx}
\usepackage{soul}
\usepackage{nicefrac}
\usepackage{booktabs}
\usepackage[super]{nth}
\usepackage{float}
\usepackage{caption}
\usepackage{subcaption}
\usepackage{dblfloatfix}
\usepackage{xcolor}
\usepackage{graphicx, import, color}

\usetikzlibrary{arrows.meta, 
	bending, 
	calc, chains, 
	decorations.pathmorphing,
	positioning
}

\usepackage{pgfplots}

\usepgfplotslibrary{fillbetween}
\usetikzlibrary{intersections}

\begin{document}

\begin{frontmatter}
	\title{Block Encoding of Sparse Matrices via Coherent Permutation}
    
    % Author affiliations
    \author[a,b]{Abhishek Setty\corref{mycorrespondingauthor}}
    \cortext[mycorrespondingauthor]{Corresponding author}
    \ead{a.setty@fz-juelich.de}  % Email of the corresponding author
    
    % Address affiliations
    \address[a]{Forschungszentrum Jülich, Institute of Quantum Control (PGI-8), D-52425 Jülich, Germany}
    \address[b]{Institute for Theoretical Physics, University of Cologne, D-50937 Cologne, Germany}
    
\begin{abstract}	
Block encoding of sparse matrices underpins quantum algorithms such as quantum singular value transformation, Hamiltonian simulation, and quantum linear system solvers, yet its efficient gate-level realization remains challenging, with index-mapping oracles constituting one important source of overhead. We introduce a block-encoding framework that focuses on the index mapping component, where coherent permutation is used as the central mechanism to optimize shift, delete, and insert operations. This provides a unified treatment of index mapping and reduces local multicontrolled X control complexity through structured compression. We further connect coherent amplitude permutation to combinatorial optimization, enabling systematic assignment of control states under hardware connectivity constraints. The resulting construction supports entry-wise block encoding for general sparse matrices, with efficiency gains in structured cases. We demonstrate the approach on representative examples and provide resource analysis using IBM superconducting backends, showing significant reductions in circuit depth.
\end{abstract}

\end{frontmatter}
\thispagestyle{empty}
%\tableofcontents

\section{Introduction}\label{sec:Intro}
Block encoding has emerged as a central primitive in modern quantum algorithms, providing a systematic way to embed a matrix into a unitary operator and thereby enabling polynomial transformation of operators via Quantum Singular Value Transformation (QSVT) \cite{gilyen2019quantum,martyn2021grand}. The idea was first used implicitly in early breakthroughs such as the Harrow-Hassidim-Lloyd (HHL) algorithm for solving linear systems of equations \cite{harrow2009quantum} and Hamiltonian simulation techniques \cite{berry2015simulating,childs2017quantum,childs2012hamiltonian}, and was later formalized by Gily\'{e}n \textit{et al}. \cite{gilyen2019quantum}. Since then, block encoding has become indispensible in a wide range of domains including quantum linear algebra, optimization, machine learning and quantum chemistry \cite{brandao2017quantum,van2020quantum,babbush2018encoding}. Succinctly, block encoding is the process of embedding a given (possibly non-unitary) matrix $A$ into a larger unitary operator $U_A$ as,
\begin{equation}\label{eq:U_A_Block_Encode}
	U_A = 
	\begin{pmatrix}
		A/\alpha & *\\
		* & *
	\end{pmatrix},
\end{equation}
where $\alpha$ is a subnormalization factor ensuring $||A/\alpha||_2 \leq 1$, $*$ denotes inconsequential blocks, and $||\cdot||_2$ is the spectral norm. The factor $\alpha$ and the presence of $*$ blocks guarantee that a unitary $U_A$ exists.

Understanding the importance of block encoding has motivated extensive efforts towards resource-efficient constructions for different matrix classes. For arbitrary dense matrices, exact block-encoding schemes based on state preparation and Linear Combination of Unitaries (LCU) techniques have been studied extensively, together with their associated resource requirements \cite{clader2023quantum,chakraborty2018power}. More recently, approximate hardware-oriented constructions based on decompositions into single- and two-qubit gates have been proposed, including the FABLE framework \cite{camps2022fable,kuklinski2024s} and subsequent improvements employing demultiplexor techniques \cite{li2025binary}. For sparse matrices, block encodings have typically been formulated in terms of black-box oracles \cite{gilyen2019quantum}, but these works often omit explicit circuit-level implementations. More detailed realizations have been provided for structured sparsity \cite{camps2024explicit}. 

The subnormalization factor $\alpha$ directly influences the complexity of QSVT-based algorithms, with the optimal choice being spectral norm $\alpha = ||A||_2$. Existing sparse block-encoding constructions \cite{sunderhauf2024block,yang2024block} generally employ larger problem-dependent factors $\alpha > ||A||_2$. Whether one can construct block encodings with subnormalization factor $\alpha = ||A||_2$ remains an important open question. 

Block encoding of sparse matrices consists predominantly of two components: data embedding via state preparation and data distribution via index-mapping operations. In this work, we focus primarily on the index-mapping component. We introduce coherent permutation operators as a central mechanism, enabling a unified treatment of shift, delete, and insert operations while reducing local multicontrolled X (MCX) control complexity through structured compression. The resulting framework supports entry-wise construction to general sparse matrices, with efficiency gains expected in structured instances. We further establish a connection between coherent amplitude permutation and combinatorial optimization, enabling systematic assignment of control states and hardware-aware circuit constructions under connectivity constraints. This perspective is particularly relevant for architectures with limited qubit connectivity \cite{linke2017experimental,beals2013efficient,kutin2007computation}. More broadly, the framework connects quantum circuit synthesis with classical optimization techniques, contributing to circuit compilation and routing strategies \cite{shende2005synthesis,amy2013meet,cowtan2019qubit,murali2019noise}. We demonstrate the approach on representative examples and provide hardware resource analysis on IBM superconducting backends, showing reductions in circuit depth. Finally, we demonstrate these ideas on representative sparse matrices with varying degrees of structure, illustrating how the framework translates abstract oracle constructions into executable circuits suitable for quantum algorithms. 

\section{Notations and Preprocessing}
\begin{figure}[t]
	\centering
	\begin{subfigure}[t]{0.25\columnwidth}
		%\centering
		\includegraphics[width=\columnwidth]{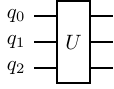}
		\caption{}
		\label{fig:1a_Qubit_ordering}
	\end{subfigure}
	\hspace{0.5cm}
	\begin{subfigure}[t]{0.4\columnwidth}
		%\centering
		\includegraphics[width=\columnwidth]{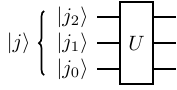}
		\caption{}
		\label{fig:1b_Qubit_ordering}
	\end{subfigure}
	\caption{\subref{fig:1a_Qubit_ordering} Qubits of circuit $U$ are numbered in increasing order from top to bottom. \subref{fig:1b_Qubit_ordering} The matrix column register $|j\rangle = |j_{n-1}\cdots j_0\rangle$. The binary representation of an integer state $|j\rangle$ is assigned to qubits in decreasing order from top to bottom.}
	\label{fig:1_Qubit_ordering}
\end{figure}

Assuming familiarity with standard conventions in the quantum computing literature, we establish following notations and conventions: 
\begin{itemize}
	\item For an $N \times N$ matrix, the $j^{\text{th}}$ column is denoted by $|j\rangle$, where $j \in [0, N-1]$. Its binary representation is given by,
	\begin{equation}\label{eq:Binary_rep}
		j \mapsto j_{n-1} \times 2^{n-1}+\dots + j_0 \times 2^0 \mapsto |j_{n-1} \dots j_1 j_0\rangle,
	\end{equation}
	where $j_k \in \{0, 1\}, k \in [0, n-1]$ and $n$ is the number of qubits.
	
	\item Qubits in a circuit diagram are ordered increasingly from top to bottom, as illustrated for the three-qubit circuit $U$ in \cref{fig:1a_Qubit_ordering}. The binary state $|j_{n-1} \cdots j_1 j_0\rangle$ is mapped to the quantum register $|q_0 q_1 \cdots q_{n-1}\rangle$, such that the highest-index qubit $|q_{n-1}\rangle$ corresponds to $|j_0\rangle$, and the lowest-index qubit $|q_0\rangle$ corresponds to $|j_{n-1}\rangle$ \cref{fig:1b_Qubit_ordering}.
	
	\item The Hamming distance \cite{hamming1950error,li2022quantum} between two binary strings $\vec{x} = (x_0, x_1, \dots, x_{n-1})^T$ and $\vec{y} = (y_0, y_1, \dots, y_{n-1})^T$, where $x_k, y_k \in \{0, 1\}$, is defined as
	\begin{equation}\label{eq:Hamming}
		D_H = |\vec{x} - \vec{y}| = \sum_{k=0}^{n-1} (x_k \oplus y_k),
	\end{equation}
	where $\oplus$ denotes the XOR operation. For instance, $D_H(010, 001) = 2$.
	
	\item Multi-controlled NOT gates are denoted by $C_{|\text{control}\rangle}^{|\text{state}\rangle}X_{|\text{target}\rangle}$, where $|\text{control}\rangle$ denote control qubits, $|\text{state}\rangle$ denote control state \cref{eq:Binary_rep}, and $X_{|\text{target}\rangle}$ denote NOT gate applied on $|\text{target}\rangle$ qubit. When convenient, we write
	\[
	C_{|\mathrm{control}_1\rangle}^{|\mathrm{state}_1\rangle}
	C_{|\mathrm{control}_2\rangle}^{|\mathrm{state}_2\rangle}
	X_{|\mathrm{target}\rangle}
	\]
	to indicate a partition of the controls into two groups while still representing a single generalized MCX gate.
\end{itemize}

The main objective of block encoding is to seek a unitary $U_A$ such that,
\begin{equation}
	(\langle 0|_{\text{anc}} \otimes \langle i |) U_A (|0\rangle_{\text{anc}} \otimes |j\rangle) = \frac{A_{ij}}{\alpha}.
\end{equation}
where $|i\rangle$ and $|j\rangle$ denote row and column registers, respectively. Also, $|0\rangle_{\text{anc}}$ denote ancilla qubits. 

We now discuss the data preprocessing needed for block encoding a sparse complex matrix $A \in \mathbb{C}^{2^n \times 2^n}$. It is important to note that this method embeds each data element only once in a row/column. Therefore, we collect unique data elements with respect to a fixed diagonal offset $d = i-j$. In the present work, two entries on the same diagonal are considered identical if they are exactly equal as complex numbers. For numerical implementations involving floating-point values, one may alternatively adopt a tolerance parameter $\varepsilon>0$ and identify entries satisfying $|A_{ij}-A_{i'j'}|<\varepsilon$ as belonging to the same equivalence class. Such tolerance-based grouping introduces an approximation to the original matrix.

For a square matrix, we define $d>0$ as lower diagonals and $d<0$ as upper diagonals. For each diagonal offset $d \in Diag = \{-(2^n-1), \dots, (2^n-1)\}$, let the nonzero entries be indexed by the local index $l = 1, \dots, |S_d|$:
\begin{equation}\label{eq:unique_set}
	\begin{split}
			S_d &= \{A_{ij} \in \mathbb{C} \; | \; A_{ij} \neq 0, i-j = d\}\\
			 &= \{s_{d,l} \in \mathbb{C} \; | \; s_{d,l} \neq 0, l = 1, \dots, |S_d|\}.
	\end{split}
\end{equation}
The associated row set is denoted by $S_r(s_{d,l})$, and the corresponding column index is given by $j=S_r(s_{d,l})-d$. 

We introduce the following labeled contribution index set:
\begin{equation}
	\begin{split}
			\mathcal{I} = \left \{(d,l,c) | \right.&d \in \text{Diag}, l = 1, \dots, |S_d|,\\
			 &c \in \{R, I\}, \gamma_c(s_{d, l}) \neq 0\left.\right\} ,
	\end{split}
\end{equation}
where
\[
\gamma_R(s_{d,l}) = \text{Re}(s_{d,l}), \quad \gamma_{I}(s_{d,l}) = \text{Im}(s_{d,l}).
\]
Here, $c=R$ and $c=I$ denote the real and imaginary components, respectively. Importantly, the labels distinguish different contributions even when their numerical values are equal.

Choose a fixed ordering of the labeled index set $\mathcal{I}$, and denote its ordered enumeration by
\begin{equation}
	(\iota_1, \dots, \iota_N), \quad N = |\mathcal{I}|,
\end{equation}
where the ordering is induced by fixed orderings of $d, l,$ and $c$. For each $\iota_k = (d(k), l(k), c(k)), \; k = 1, \dots, N$, define
\begin{equation}
	m_k = \begin{cases}
		|\text{Re}(s_{d(k), l(k)})|, \quad c(k) = \mathrm{R},\\
		|\text{Im}(s_{d(k), l(k)})|, \quad c(k) = \mathrm{I},\\
	\end{cases}
\end{equation}
and
\begin{equation}
	p_k = \begin{cases}
		\text{sgn}(\text{Re}(s_{d(k), l(k)})), \quad c(k) = \mathrm{R},\\
		\text{sgn}(\text{Im}(s_{d(k), l(k)}))\mathrm{i}, \quad c(k) = \mathrm{I},\\
	\end{cases}
\end{equation}
where $\text{sgn}(a) = \frac{a}{|a|}, \; a \in \mathbb{R} \backslash \{0\}$. The data and sign vectors are then defined by
\begin{equation}\label{eq:v_data}
	v_{\text{data}} = (m_1, \dots, m_N),
\end{equation}
\begin{equation}\label{eq:v_sign}
	 v_{\text{sign}} = (p_1, \dots, p_N).
\end{equation}
By construction, all entries of $v_{\text{data}}$ are non-negative real numbers obtained from the magnitudes of nonzero real or imaginary components. The dimension of the data vector is $\text{dim}(v_{\text{data}})=N$. Each $k^{\text{th}}$ element in the flattened vectors has the unique back-mapping $k \mapsto \iota_k = (d(k), l(k), c(k))$, where $d(k)$ denotes the diagonal offset, $l(k)$ denotes the local position within $S_{d(k)}$, and $c(k)$ identifies the real or imaginary component. We therefore define $d_k = d(k)$, and $S_r(k) = S_r(s_{d(k), l(k)})$. The corresponding column index is then $j_k = S_r(k) - d_k$.

We summarize the notations of data structures and quantum registers as in \cref{table:data_structure}.
\begin{table}[ht]
	\centering
	\begin{tabular}{|c|c|}
		\hline
		\multicolumn{2}{|c|}{Data structure and registers}\\
		\hline
		Row register & $|i\rangle$ \\
		Column register & $|j\rangle$\\
		Value of $k^{\text{th}}$ data element & $v_{\text{data},k}$\\
		Sign of $k^{\text{th}}$ data element & $v_{\text{sign},k}$\\
		Data register basis state & $|k\rangle_{\text{data}}$\\
		Diagonal shift & $d_{k}$\\
		Row set & $S_r(k)$\\
		Delete flag & $|b\rangle_{\text{del}}$\\
		\hline
	\end{tabular}
	\caption{Notations of data structures and quantum registers used in block encoding.}
	\label{table:data_structure}
\end{table}

\section{Block Encoding}\label{sec:block_encoding} 
In this section, we elaborate the PREP/UNPREP-based block encoding \cite{sunderhauf2024block,yang2024block} and present a framework for constructing the corresponding quantum circuits. Primarily, block encoding consists of two oracles, namely, state preparation and index mapping. 

Let the number of data qubits $m= \left\lceil \log_2 \textnormal{dim}(v_{\textnormal{data}}) \right\rceil$ for the collected data $v_{\text{data}}$ and signs $v_{\text{sign}}$ (see \cref{eq:v_data,eq:v_sign}). Then the state preparation oracles PREP (\cref{eq:PREP}) and UNPREP (\cref{eq:UNPREP}) together embed $v_{\text{sign}} \circ v_{\text{data}}$ into the amplitudes of a quantum state. Here $\circ$ denote element wise multiplication between two vectors. 

Note that the state vectors are padded with zeros to fill the $2^m$ basis states required for state preparation. These additional basis states do not correspond to physical data elements but serve only as padding. Since the phase associated with a zero amplitude is physically irrelevant, we adopt the convention $\operatorname{sgn}(0)=0$ to obtain a canonical representation, which is convenient for the compression procedures introduced later.
\begin{equation}
	\resizebox{\columnwidth}{!}{$
		\begin{split}\label{eq:PREP}
			&\textnormal{PREP} |0\rangle_{\textnormal{data}}^{\otimes m} = \frac{1}{\sqrt{\sum_{k=0}^{\textnormal{dim}(v_{\textnormal{data}})-1} v_{\textnormal{data},k}}}\\
			&\left( \sum_{k=0}^{\textnormal{dim}(v_{\textnormal{data}})-1} v_{\textnormal{sign},k} \sqrt{v_{\textnormal{data},k}} |k\rangle_{\textnormal{data}} + \sum_{k=\textnormal{dim}(v_{\textnormal{data}})}^{2^m-1} 0 |k\rangle_{\textnormal{data}} \right),
		\end{split}
		$}
\end{equation}
\begin{equation}
	\resizebox{\columnwidth}{!}{$
		\begin{split}\label{eq:UNPREP}
			&\textnormal{UNPREP}^{\dag} |0\rangle_{\textnormal{data}}^{\otimes m} = \frac{1}{\sqrt{\sum_{k=0}^{\textnormal{dim}(v_{\textnormal{data}})-1} v_{\textnormal{data},k}}}\\
			&\left( \sum_{k=0}^{\textnormal{dim}(v_{\textnormal{data}})-1} \sqrt{v_{\textnormal{data},k}} |k\rangle_{\textnormal{data}} + \sum_{k=\textnormal{dim}(v_{\textnormal{data}})}^{2^m-1} 0 |k\rangle_{\textnormal{data}} \right),
		\end{split}
		$}
\end{equation}
where the left projection $\langle0|_{\textnormal{data}}^{\otimes m} \textnormal{UNPREP}$ is determined by $\textnormal{UNPREP}^{\dag} |0\rangle_{\textnormal{data}}^{\otimes m}$.

The index mapping oracle is composed of two oracles such as shift $O_{\text{shift}}$ and delete $O_{\text{del}}$. The shift oracle performs the injective mapping
\begin{equation}
f_k(j): j \mapsto i = \operatorname{mod}(j+d_k,2^n),
\end{equation}
thereby mapping column indices $j$ to row indices $i$
along the diagonal offset $d_k=i-j$. The oracle $O_{\text{del}}$ performs deletion of elements from rows $f_k(j) \notin S_r(k)$.

\begin{figure}[t]
	\centering
	\includegraphics[width=\columnwidth]{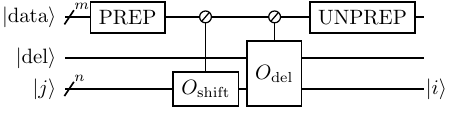}
	\captionsetup{width=\columnwidth}
	\caption{Circuit structure for block encoding of sparse matrices. A single qubit is represented by a plain wire (del qubit), while multiple qubits are depicted as a wire marked with a short slash, labeled with the number of qubits it contains ($m$ for data qubits and $n$ for matrix qubits $|j\rangle$). The PREP and UNPREP blocks denote state-preparation operators. The oracles $O_{\text{shift}}$ and $O_{\text{del}}$ uses multi-controlled gates denoted by circle with a slash. The control values are determined by the respective state in binary representation \cref{eq:Binary_rep}, following the qubit ordering convention shown in \cref{fig:1_Qubit_ordering}.} 
	\label{fig:2_BE}
\end{figure}

With these oracles in place, the block encoding scheme is formulated as described in \cref{theorem_1}.
\begin{theorem} \label{theorem_1}
	Let $A \in \mathbb{C}^{2^n \times 2^n}$ be a matrix that has collected data $v_{\textnormal{data}}$ and sign $v_{\textnormal{sign}}$. If there exists shift oracle $O_{\textnormal{shift}}$ such that
	\begin{equation}
		\begin{split}
					O_{\textnormal{shift}} &| k \rangle_{\textnormal{data}} |b\rangle_{\textnormal{del}} |j\rangle = | k \rangle_{\textnormal{data}} |b\rangle_{\textnormal{del}} |f_k(j)\rangle \\
					&= | k \rangle_{\textnormal{data}} |b\rangle_{\textnormal{del}} | \textnormal{mod}(j + d_k, 2^n)\rangle,
		\end{split}
	\end{equation}
and delete oracle $O_{\textnormal{del}}$ such that,
\begin{equation}
		\begin{split}
O_{\textnormal{del}} & | k \rangle_{\textnormal{data}} |b\rangle_{\textnormal{del}} |f_k(j)\rangle = \\
	&| k \rangle_{\textnormal{data}} |b \oplus \chi_k(f_k(j)) \rangle_{\textnormal{del}} |f_k(j)\rangle,\\
\end{split}
\end{equation}
where 
\[
\chi_k(f_k(j)) =
\begin{cases}
	0 \; \text{if} \; f_k(j) \in S_r(k)\\
	1 \; \text{else}
\end{cases}
\]
and two state preparation oracles \textnormal{PREP} and \textnormal{UNPREP}. Then the unitary, $U_A = (\textnormal{UNPREP} \otimes I_{2^{n+1}}) O_{\textnormal{del}} O_{\textnormal{shift}} (\textnormal{PREP} \otimes I_{2^{n+1}})$, as shown in \cref{fig:2_BE}, can block encode $A$ with the subnormalization $\alpha = \sum_{k}v_{\textnormal{data},k}$.
\end{theorem}

\textit{Proof:} To recover the matrix from its block encoding, the ancilla qubits (i.e., the data and delete qubits) are initialized and postselected in the state $|0\rangle$. This is achieved by initializing the bottom register with $|j\rangle$ and postselecting (or measuring) the outcome $|i\rangle$, as follows:
\[
\begin{split}
&
\langle 0 |_{\text{data}}^{\otimes m} \langle 0|_{\text{del}} \langle i| \; U_A \; |0\rangle_{\text{data}}^{\otimes m} |0\rangle_{\text{del}} |j\rangle \\
&
\begin{split}
&= \frac{1}{\sqrt{\sum_{k'}v_{\text{data},k'}\sum_{k}v_{\text{data},k}}} \sum_{k',k} \langle k' |_{\text{data}} \langle 0|_{\text{del}} \langle i|\\
& v_{\text{sign},k}  \sqrt{v_{\text{data},k'}v_{\text{data},k}} \; O_{\text{del}} O_{\text{shift}} |k\rangle_{\text{data}} |0\rangle_{\text{del}} |j\rangle
\end{split}\\
&
\begin{split}
&= \frac{1}{\sqrt{\sum_{k'}v_{\text{data},k'}\sum_{k}v_{\text{data},k}}} \sum_{k',k} \langle k' |_{\text{data}} \langle 0|_{\text{del}} \langle i|\\
	& v_{\text{sign},k}  \sqrt{v_{\text{data},k'}v_{\text{data},k}} \; O_{\text{del}} |k\rangle_{\text{data}} |0\rangle_{\text{del}} |f_k(j)\rangle
\end{split}\\
&
\begin{split}
&= \frac{1}{\sqrt{\sum_{k'}v_{\text{data},k'}\sum_{k}v_{\text{data},k}}} \sum_{k',k} \langle k' |_{\text{data}} \langle 0|_{\text{del}} \langle i|\\
& v_{\text{sign},k}  \sqrt{v_{\text{data},k'}v_{\text{data},k}} \; \mathbf{1}_{S_r(k)}(i)  \; \delta_{k,k'} \; \delta_{i, f_k(j)}\\
& \hspace{0.5cm} |k\rangle_{\text{data}} |0\rangle_{\text{del}} |f_k(j)\rangle
\end{split}\\
& = \frac{1}{\alpha} \sum_{\substack{k: f_k(j)=i \\ i \in S_r(k)}}v_{\text{sign},k} v_{\text{data},k},
\end{split}
\]
where $\delta$ denotes a kronecker delta function $\delta_{ij} = 1 \text{ if } i = j, \text{else } 0$ and indicator function $\mathbf{1}_{S_r(k)}(i) = 1$ if $i \in S_r(k)$, else $0$. 

In case of complex entries, we add its real and imaginary components within the block encoded matrix. Consider $k^{\text{th}}$ and $l^{\text{th}}$ data elements in $v_{\text{data}}$ and $v_{\text{sign}}$ correspond to $\text{Re}(A_{ij})$ and $\text{Im}(A_{ij})$, respectively. Then the above proof can block encode a complex entry as follows,
\begin{equation}\label{eq:complex_addition}
	\begin{split}
			\frac{A_{ij}}{\alpha} &= \frac{1}{\alpha} (\text{Re}(A_{ij}) + \text{Im}(A_{ij})\mathrm{i})\\
			&= \frac{1}{\alpha} (v_{\text{sign},k} v_{\text{data},k} + v_{\text{sign},l} v_{\text{data},l}).
	\end{split}
\end{equation}
\begin{figure*}[t]
	\centering
	\begin{subfigure}[t]{0.27\textwidth}
		%\centering
		\includegraphics[width=\textwidth]{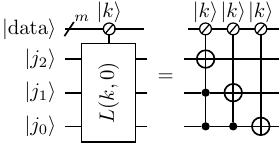}
		\caption{} 
		\label{fig:3a_Shifting}
	\end{subfigure}
	\hfill
	\begin{subfigure}[t]{0.27\textwidth}
		%\centering
		\includegraphics[width=\textwidth]{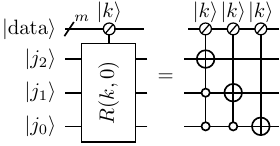}
		\caption{}
		\label{fig:3b_Shifting}
	\end{subfigure}
	\hfill
	\begin{subfigure}[t]{0.27\textwidth}
		%\centering
		\includegraphics[width=\textwidth]{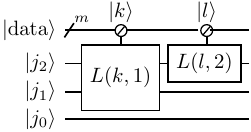}
		\caption{}
		\label{fig:3c_Shifting}
	\end{subfigure}
	\hfill
	\begin{subfigure}[t]{0.14\textwidth}
		%\centering
		\includegraphics[width=\textwidth]{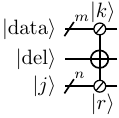}
		\caption{}
		\label{fig:3d_Deletion}
	\end{subfigure}
	\caption{\subref{fig:3a_Shifting} Circuit illustration of left shift oracle $L(k, 0)$ (refer \cref{eq:left_shift} with three matrix qubits $|j\rangle$ and $m$ data qubits. Here the control state is given by $|k\rangle$ and denoted by circle with a slash. \subref{fig:3b_Shifting} Similarly, circuit illustration of right shift oracle $R(k, 0)$ (refer \cref{eq:right_shift}). \subref{fig:3c_Shifting} Left shift oracle of $k^{\text{th}}$ and $l^{\text{th}}$ data elements by two and four columns, respectively \subref{fig:3d_Deletion} Delete oracle $O_{\text{del}}$ illustrating deletion of $k^{\text{th}}$ data element from $r^{\text{th}}$ row (refer \cref{eq:O_del}).}
	\label{fig:3_Index_Mapping}
\end{figure*}
\subsection{State Preparation Oracle}\label{sec:State_prep_oracle}
In this section, we briefly discuss the existing methods of state preparation oracles PREP/UNPREP, whose task is to embed data into the amplitudes of a quantum state. Möttönen et al. \cite{mottonen2004transformation, mottonen2004quantum} introduced a state preparation method based on uniformly controlled rotation gates, which can be decomposed into either multi-controlled rotations or sequences of single- and two-qubit gates. This construction leverages classical preprocessing—such as Gray code ordering—to structure the circuit efficiently, reducing the number of controlled rotations required. Later, Iten et al. \cite{iten2016quantum} proposed a hardware-oriented approach that decomposes arbitrary isometries exactly into single-qubit and CNOT gates via a recursive synthesis procedure based on the cosine-sine decomposition. Both methods are exact and require no ancilla qubits, but generally scale exponentially in gate count and depth for arbitrary state preparation. More recent work has explored depth-optimized schemes \cite{zhang2022quantum}, achieving circuit depths of $\mathcal{O}(m)$ for an $m$-qubit state or $\mathcal{O}(\log_2(m s))$ for $s$-sparse states. These improvements, however, often come at the cost of introducing additional ancilla qubits, which in some cases can scale exponentially requiring $\mathcal{O}(2^m)$ ancilla qubits.

\subsection{Index Mapping Oracle}\label{sec:Index_mapping_oracle}
Without the index mapping oracles $O_{\text{shift}}$ and $O_{\text{del}}$ in \cref{fig:2_BE}, the block-encoded matrix reduces to a diagonal form (refer \cref{theorem_1}),
\begin{equation}\label{eq:diagonal_block_encoded}
	\resizebox{\columnwidth}{!}{$
	A/\alpha =
\begingroup
\setlength{\arraycolsep}{2pt}
\renewcommand{\arraystretch}{0.7} \begin{bmatrix}
	x &\\
	& \ddots &\\
	&& x
\end{bmatrix}, x = \frac{\sum_{k}v_{\text{sign,k}} v_{\text{data},k}}{\sum_k v_{\text{data},k}}.
\endgroup
$}
\end{equation}

The oracle $O_{\mathrm{shift}}$ \cite{camps2024explicit} may equivalently be interpreted as fixing the row index and updating the column register. Under the convention $d_k=i-j, j=i-d_k,$ and therefore a main-diagonal entry $(i,i)$ is mapped to
\[
(i,i)\mapsto
(i,(i-d_k)\bmod 2^n).
\]
Specifically, we define the shift oracle $O_{\text{shift}}$ as,
\begin{equation}\label{eq:O_shift}
	O_{\text{shift}}(k, d) = \begin{cases}
		\prod_{b\in B(|d|)}L(k, b) \; \text{if} \; d>0,\\
		I \; \text{if} \; d=0,\\
		\prod_{b\in B(|d|)}R(k, b) \; \text{if} \; d<0,\\
	\end{cases}
\end{equation}
where we define absolute value of diagonal $|d|$ in binary form:
\begin{equation}
	|d| = \sum_{l=0}^{n-1}b_l 2^l, b_l \in \{0,1\}.
\end{equation}
For an integer $|d|$ in binary $(b_{n-1}\cdots b_0)_2$, we define a set of 1-bit positions  as $B(|d|) = \{l | b_l=1\}$. In other words, a shift $|d|$ is decomposed into a sequence of conditional shifts by powers of two. Then for $d>0$, we define a left shift oracle for $k^{\text{th}}$ data element $v_{\text{data},k}$ shifting left by $2^b$ columns as:
\begin{equation}\label{eq:left_shift}
	L(k, b) = \prod_{l=b}^{n-1}C_{|\text{data}\rangle}^{|k\rangle} C_{|j_{l-1}\cdots j_b\rangle}^{|1\rangle^{\otimes (l-b)}} X_{|j_l\rangle}.
\end{equation}
Note that the order of operators are written from left to right and the right most operator is applied first in the circuit. Similarly, for $d<0$, we define a right shift oracle for $k^{\text{th}}$ data element $v_{\text{data},k}$ shifting right by $2^b$ columns as:
\begin{equation}\label{eq:right_shift}
	R(k, b) = \prod_{l=b}^{n-1}C_{|\text{data}\rangle}^{|k\rangle} C_{|j_{l-1}\cdots j_b\rangle}^{|0\rangle^{\otimes (l-b)}} X_{|j_l\rangle}.
\end{equation}

To visualize the circuit, we consider three matrix qubits $|j\rangle$ and represent left shift (d=1) of $k^{\text{th}}$ data element by one column $L(k, 0)$ (see \cref{eq:left_shift}) in \cref{fig:3a_Shifting}. Similarly, the right shift (d=-1) of $k^{\text{th}}$ data element by one column $R(k, 0)$ (see \cref{eq:right_shift}) is shown in \cref{fig:3b_Shifting}. Furthermore, the representation of shifting left the $k^{\text{th}}$ and $l^{\text{th}}$ data elements by two and four columns, respectively is shown in \cref{fig:3c_Shifting}. 

To visualize the block encoded matrix after shifting, we represent an example as follows. Consider the data vector $v_{\text{data}} = [\psi_0, \psi_1, \psi_2, \psi_3, \psi_4]^T$, sign vector $v_{\text{sign}}=[1, 1, -\mathrm{i}, -1, \mathrm{i}]^T$ and block encoding matrix to be $8 \times 8$ requiring three matrix qubits. If we apply the operator $L(4,2)L(3, 1)L(3, 0)L(2, 1)L(1, 1)L(0, 0)$ (see \cref{eq:left_shift,eq:O_shift} and \cref{fig:3a_Shifting}), then the block encoded matrix will be given as
\begin{equation}\label{eq:L_Shift}
		\resizebox{\columnwidth}{!}{$
	\begin{split}
	&A/\alpha = \frac{1}{\psi_0 + \psi_1 + \psi_2 + \psi_3 + \psi_4}\\
	&\begingroup
	\setlength{\arraycolsep}{0 pt}
	\renewcommand{\arraystretch}{1}
\begin{bmatrix} 
	0 & 0 & 0 & 0 & \psi_4\mathrm{i} & -\psi_3 & \psi_1-\psi_2\mathrm{i} & \psi_0\\
	\psi_0 & 0 & 0 & 0 & 0& \psi_4\mathrm{i} & -\psi_3 &\psi_1-\psi_2\mathrm{i}\\
	\psi_1-\psi_2\mathrm{i} & \psi_0 & 0 & 0 & 0 & 0 & \psi_4\mathrm{i} & -\psi_3\\
	-\psi_3 & \psi_1-\psi_2\mathrm{i} &\psi_0 & 0 & 0 & 0 & 0 & \psi_4\mathrm{i}\\
	\psi_4\mathrm{i} & -\psi_3 & \psi_1-\psi_2\mathrm{i} & \psi_0 & 0 & 0 & 0 & 0\\
	0 & \psi_4\mathrm{i} & -\psi_3 & \psi_1-\psi_2\mathrm{i} & \psi_0 & 0 & 0 & 0\\
	0 & 0 & \psi_4\mathrm{i} & -\psi_3 & \psi_1-\psi_2\mathrm{i} & \psi_0 & 0 & 0\\
	0 & 0 & 0 & \psi_4\mathrm{i} & -\psi_3 & \psi_1-\psi_2\mathrm{i} & \psi_0 & 0\\
\end{bmatrix} 
\endgroup
	\end{split}
	$}
\end{equation}
Note that complex entries are block encoded by adding its non-zero real and imaginary components (see \cref{eq:complex_addition}). Similarly, if we apply the operator $R(4,2)R(3, 1)R(3, 0)R(2, 1)R(1, 1)R(0, 0)$ (see \cref{eq:right_shift,eq:O_shift} and \cref{fig:3b_Shifting}), then the block encoded matrix will be given as
\begin{equation}\label{eq:R_Shift}
	\resizebox{\columnwidth}{!}{$
	\begin{split}
	&A/\alpha = \frac{1}{\psi_0 + \psi_1 + \psi_2 + \psi_3 + \psi_4}\\
	&\hspace{-0.3cm}\begingroup
	\setlength{\arraycolsep}{0 pt}
	\renewcommand{\arraystretch}{1}
	\begin{bmatrix}
		0 & \psi_0 & \psi_1 - \psi_2\mathrm{i} & -\psi_3 & \psi_4\mathrm{i} & 0 & 0 & 0\\
		0 & 0 & \psi_0 & \psi_1 - \psi_2\mathrm{i} & -\psi_3& \psi_4\mathrm{i} & 0 & 0\\
		0 & 0 & 0 & \psi_0 & \psi_1 - \psi_2\mathrm{i} & -\psi_3 & \psi_4\mathrm{i} & 0\\
		0 & 0 & 0 & 0 & \psi_0 & \psi_1 - \psi_2\mathrm{i} & -\psi_3 & \psi_4\mathrm{i}\\
		\psi_4\mathrm{i} & 0 & 0 & 0 & 0 & \psi_0 & \psi_1 - \psi_2\mathrm{i} & -\psi_3\\
		-\psi_3 & \psi_4\mathrm{i} & 0 & 0 & 0 & 0 & \psi_0 & \psi_1 - \psi_2\mathrm{i}\\
		\psi_1 - \psi_2\mathrm{i} & -\psi_3 & \psi_4\mathrm{i} & 0 & 0 & 0 & 0 & \psi_0\\
		\psi_0 & \psi_1 - \psi_2\mathrm{i} & -\psi_3 & \psi_4\mathrm{i} & 0 & 0 & 0 & 0\\
	\end{bmatrix} 
	\endgroup
	\end{split}
	$}
\end{equation}

The delete oracle \cite{sunderhauf2024block} for removing the $k^{\text{th}}$ data element from the $r^{\text{th}}$ row is defined as:
\begin{equation}\label{eq:O_del}
	O_{\text{del}}(k, r) = D^{|k\rangle}_{|r\rangle} = C_{|\text{data}\rangle}^{|k\rangle}C_{|j\rangle}^{|r\rangle}X_{|\text{del}\rangle}.
\end{equation}
The corresponding gate is illustrated in \cref{fig:3d_Deletion}. 

To insert a single data element into one or a few rows, one could in principle delete the data element from all remaining rows. However, this approach may require an exponential number of MCX gates in general. To avoid this overhead, we instead introduce an insert operator $I^{|k\rangle}_{|r\rangle}$, which directly places the $k^{\text{th}}$ data element into the $r^{\text{th}}$ row without requiring exponential MCX gates. The construction of this operator is detailed later in \cref{sec:Insert}.

\section{Composition of Multi-Controlled X Gates}\label{sec:MCX_Composition}
In this section, we analyze the composition and simplification of multi-controlled $X$ (MCX) gates that arise in index-mapping oracles.

\begin{figure}[t]
	\centering
	\includegraphics[width=\columnwidth]{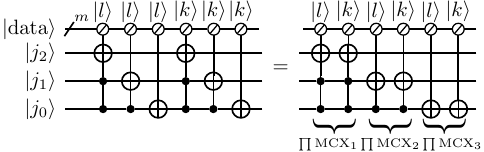}
	\caption{Consecutive left shift oracles $L(k, 0)L(l, 0)$ on three matrix qubits $|j\rangle$. The MCX gates can be grouped into three compositions $\prod \text{MCX}_i$ (see \cref{eq:MCX_composition}).}
	\label{fig:4_Commute}
\end{figure}

Consider two successive shift operators $L(k,0)L(l,0)$ (see \cref{eq:left_shift,fig:3a_Shifting}) acting on three matrix qubits as shown in \cref{fig:4_Commute}. Expanding both operators, we obtain
\begin{equation}\label{eq:MCX_composition}
	\resizebox{\columnwidth}{!}{$
		\begin{split}
			&L(k,0)L(l,0)
			=
			\biggl[\left(C_{|\text{data}\rangle}^{|k\rangle}X_{|j_0\rangle}\right)
			\left(C_{|\text{data}\rangle}^{|k\rangle}C_{|j_0\rangle}^{|1\rangle}X_{|j_1\rangle}\right)\biggr.\\
			&\left(C_{|\text{data}\rangle}^{|k\rangle}C_{|j_1j_0\rangle}^{|1\rangle^{\otimes 2}}X_{|j_2\rangle}\right)\biggr] 
			\biggl[ \left(C_{|\text{data}\rangle}^{|l\rangle}X_{|j_0\rangle}\right)
			\left(C_{|\text{data}\rangle}^{|l\rangle}C_{|j_0\rangle}^{|1\rangle}X_{|j_1\rangle}\right)\biggr.\\
			&\biggl. \left(C_{|\text{data}\rangle}^{|l\rangle}C_{|j_1j_0\rangle}^{|1\rangle^{\otimes 2}}X_{|j_2\rangle}\right)\biggr],\\
			&= \biggl[ \left(C_{|\text{data}\rangle}^{|k\rangle}X_{|j_0\rangle}\right)
			\left(C_{|\text{data}\rangle}^{|l\rangle}X_{|j_0\rangle}\right)\biggr] \\ &\hspace{0.45cm} \biggl[\left(C_{|\text{data}\rangle}^{|k\rangle}C_{|j_0\rangle}^{|1\rangle} X_{|j_1\rangle} \right)
			\left(C_{|\text{data}\rangle}^{|l\rangle}C_{|j_0\rangle}^{|1\rangle}X_{|j_1\rangle} \right)\biggr] \\ &\hspace{0.45cm} \biggl[\left(C_{|\text{data}\rangle}^{|k\rangle}C_{|j_1j_0\rangle}^{|1\rangle^{\otimes 2}}X_{|j_2\rangle} \right)
			\left(C_{|\text{data}\rangle}^{|l\rangle}C_{|j_1j_0\rangle}^{|1\rangle^{\otimes 2}}X_{|j_2\rangle} \right)\biggr],\\
			&= \bigg[\prod \text{MCX}_1\bigg]\bigg[\prod \text{MCX}_2\bigg]\bigg[\prod \text{MCX}_3\bigg].
		\end{split}
		$}
\end{equation}

Since the two operators act on the same qubits, we can reorder the factors by grouping terms acting on identical control and target qubits. This reordering is valid because the controlled operations corresponding to distinct control states $|k\rangle$ and $|l\rangle$ act on orthogonal subspaces, i.e., $|k\rangle\langle k| |l\rangle\langle l| = 0,$ which implies that the corresponding MCX gates commute. Therefore, each group forms a commuting composition of MCX gates acting on the same qubits, which we denote as $\bigg[\prod \text{MCX}_1\bigg]\bigg[\prod \text{MCX}_2\bigg]\bigg[\prod \text{MCX}_3\bigg]$ (see \cref{fig:4_Commute}). A similar structure arises in deletion operations across multiple rows (see \cref{fig:3d_Deletion}). In both cases, we obtain compositions of MCX gates acting on identical sets of control and target qubits. Such compositions can be simplified under suitable conditions.

Let $G = \{0, \dots, P-1\}$ denote the index set of $P$ qubits (see \cref{eq:Binary_rep,fig:1_Qubit_ordering}). Let
\[
S_1 = \{a^j\}_{j=0}^{2^P-1}
\]
be the set of all $P$-bit binary strings, where
\[
a^j = (a^j_i)_{i \in G}, \quad a^j_i \in \{0,1\}.
\]

Consider a composition of $2^n$ MCX gates $(1 \leq n < P)$:
\[
\prod_{j=0}^{2^n-1} C_G^{|a^j\rangle} X_{|t\rangle},
\]
where $|t\rangle$ is the target qubit and the controls act on $G$ qubits. The simplification of such compositions is possible under certain conditions as described in the following \cref{theorem:MCX_Composition}.

\begin{theorem}\label{theorem:MCX_Composition}
	Let $S_2 = \{a^j\}_{j=0}^{2^n-1} \subset S_1 = \{0,1\}^P$. Suppose there exists a subset of indices $F \subset G$ with $|F| = P-n$ such that
	\[
	a^j_i = a^k_i \quad \forall i \in F, \; \forall j,k,
	\]
	and the substrings on the remaining indices satisfy
	\[
	\{ (a^j_i)_{i \in G \setminus F} \}_{j=0}^{2^n-1} = \{0,1\}^n,
	\]
	Then
	\[
	%\resizebox{\columnwidth}{!}{$
	\prod\limits_{j=0}^{2^n-1} C_G^{|a^j\rangle} X_{|t\rangle} = C_F^{|\tilde{a}\rangle} X_{|t\rangle},
	%$}
	\]
	where $|\tilde{a}\rangle = \bigotimes_{i \in F} |a_i\rangle$.
\end{theorem}
\textit{Proof}: Consider each MCX gate is written as,
\[
C_G^{|a^j\rangle} X_{|t\rangle} = |a^j\rangle\langle a^j| \otimes X_{|t\rangle} + (I - |a^j\rangle \langle a^j|) \otimes I_{|t\rangle},
\]
where the identity operator $I$ has dimension corresponding to the number of qubits it acts on. Let $F \subset G$ denote the set of fixed indices where $a^j_i = a^k_i \; \forall i \in F, \forall j, k \in \{0, \dots, 2^n-1\}$, and let $G \setminus F$ denote the varying indices. We define
\[
|\tilde{a}\rangle = \bigotimes_{i \in F} |a_i\rangle,\quad
|a'^j\rangle = \bigotimes_{i \in G \setminus F} |a^j_i\rangle,
\]
where the substrings $\{a'^j\}_{j=0}^{2^n-1} = \{0,1\}^n$ enumerate all $2^n$ binary strings.

Without loss of generality, we assume that the qubits are ordered such that the indices in $F$ precede those in $G \setminus F$. This does not affect the final operation and allows us to rewrite the control state as,\vspace{-0.2cm}
\[
|a^j\rangle \langle a^j| = |\tilde{a}\rangle \langle \tilde{a}| \otimes |a'^j\rangle \langle a'^j|.
\]
Then the product of MCX gates can be written as,\vspace{-0.2cm}
\[
\begin{split}
\prod\limits_{j=0}^{2^n-1} C_G^{|a^j\rangle} X_{|t\rangle} = \prod\limits_{j=0}^{2^n-1} \! &\left[ |\tilde{a}\rangle \langle \tilde{a} | \! \otimes \! |a'^j\rangle \langle a'^j| \! \otimes \! X_{|t\rangle} \! + \right. \\
&\left. (I \! -\! (|\tilde{a}\rangle \langle \tilde{a} | \! \otimes \! |a'^j\rangle \langle a'^j|)) \! \otimes \! I_{|t\rangle} \right].
\end{split}
\]
In this product expansion, due to orthogonality $\langle a'^j| a'^k\rangle = \delta_{jk}$, the cross-terms get eliminated resulting in the summation as,\vspace{-0.2cm}
\[
\begin{split}
\prod\limits_{j=0}^{2^n-1} C_G^{|a^j\rangle} X_{|t\rangle} = \sum_{j=0}^{2^n-1} &\left(  |\tilde{a}\rangle \langle \tilde{a}| \otimes |a'^j\rangle \langle a'^j| \otimes X_{|t\rangle} + \right. \\
&\left. (I - |\tilde{a}\rangle \langle \tilde{a}|) \otimes |a'^j\rangle \langle a'^j| \otimes I_{|t\rangle} \right),
\end{split}
\]
where $\sum\limits_{j=0}^{2^n-1} |a'^j\rangle \langle a'^j| = I$, since the summation of projectors over complete computational basis forms the identity operator. Therefore, the above expression can be further simplified to,\vspace{-0.2cm}
\[\resizebox{\columnwidth}{!}{$
\begin{split}
\prod\limits_{j=0}^{2^n-1} C_G^{|a^j\rangle} X_{|t\rangle} &= |\tilde{a}\rangle \langle \tilde{a}| \! \otimes \! I \! \otimes \! X_{|t\rangle} \! + \! \left( I \! - \! |\tilde{a}\rangle \langle \tilde{a}| \right) \! \otimes \! I \! \otimes \! I_{|t\rangle}\\[-0.3cm]
& = C_F^{|\tilde{a}\rangle} X_{|t\rangle}.
\end{split}
$}\]
In the special case $n = P$, the control set spans the entire computational basis, and hence
\begin{equation}\label{eq:n=p}
	\prod\limits_{j=0}^{2^P-1} C_G^{|a^j\rangle} X_{|t\rangle} = X_{|t\rangle}.
\end{equation}

\section{Combinatorial Optimization Based Mapping of Basis States}\label{sec:Mapping}

In the previous section (see \cref{sec:MCX_Composition}), we showed that a composition of MCX gates can be compressed when the control set 
$S_2 = \{a^j\}_{j=0}^{2^n-1}$ exhibits a fixed set of indices $F \subset G$, with $|F| = P-n$, such that 
\[
a^j_i = a^k_i, \quad \forall i \in F, \;\forall j,k \in \{0,\dots,2^n-1\},
\]
and the substrings over the remaining indices enumerate all binary strings of length $n$. 

If $S_2$ does not satisfy these structural conditions, then the objective is to find a structured set $S_3$ together with a bijection $\phi: S_2 \mapsto S_3$ which satisfies the conditions required for compression (see \cref{theorem:MCX_Composition}).

For a given set of fixed indices $F \subset G$, with $|F| = P-n$, we define a structured set 
$S_3 = \{b^j\}_{j=0}^{2^n-1} \subset \{0,1\}^P$ such that
\[
b^j_i = b^k_i, \quad \forall i \in F, \;\forall j,k,
\]
and the substrings over the remaining indices exhaust all configurations,
\[
\{ (b^j_i)_{i \in G \setminus F} \}_{j=0}^{2^n-1} = \{0,1\}^n.
\]
This construction ensures that $|S_3| = 2^n$ and that $S_3$ satisfies the structural conditions required for MCX compression. We then formalize the mapping of basis states as a combinatorial optimization problem as follows.

\begin{theorem}\label{theorem:Mapping}
Given an arbitrary set of binary strings $S_2\subset\{0,1\}^P$ with $|S_2|=2^n$ and a target set $S_3$ constructed according to the mode-based rule, the bijection $\phi:S_2\rightarrow S_3$ that minimizes
\[
\sum_{x\in S_2}D_H(x,\phi(x))
\]
can be obtained by solving a linear assignment problem over the unmatched subsets $S_2\setminus S_3$ and $S_3\setminus S_2$, while fixing identity mapping on $S_2 \cap S_3$.
\end{theorem}

\textit{Proof}: Let $S_2=\{a^j\}_{j=0}^{2^n-1}$. The mode-based rule first constructs the target set $S_3=\{b^j\}_{j=0}^{2^n-1}$ by selecting the most frequent fixed-bit pattern \begin{equation}\label{eq:mode_bits}
	\tilde{a} =	\operatorname{mode}	\big(\{a_F^j\}_{j=0}^{2^n-1}\big),
\end{equation}
where $a_F^j$ denotes the restriction of $a^j$ to the fixed index set $F$. The target set is then defined by fixing
\[
b_i^j=\tilde{a}_i,\qquad i\in F,
\]
while allowing the remaining bits to enumerate all binary strings over
$G\setminus F$. Consequently, $|S_3|=2^n$.

Define
\[
S_2'=S_2\setminus S_3, \qquad S_3'=S_3\setminus S_2.
\]
Since $|S_2|=|S_3|$, it follows that $|S_2'|=|S_3'|$. Elements in $S_2 \cap S_3$ are assigned via identity mapping and do not contribute to the optimization.

Construct the cost matrix
\[
C_{jk}=D_H(a^j,b^k), \qquad a^j\in S_2',\; b^k\in S_3',
\]
and solve the assignment problem
\begin{equation}\label{eq:Optimization}
	\begin{aligned}
		\min_{x_{jk}}\quad & \sum_j\sum_k C_{jk}x_{jk},\\
		\textnormal{s.t.}\quad & \sum_kx_{jk}=1,\quad\forall j,\\
		& \sum_jx_{jk}=1,\quad\forall k,\\
		&x_{jk}\in\{0,1\},\quad\forall j,k.
	\end{aligned}
\end{equation}
Here $x_{jk}=1$ if and only if $\phi(a^j)=b^k$.

\cref{eq:Optimization} is the classical linear assignment problem, which admits an optimal solution computable in $O(|S_2'|^3)$ time using the Hungarian algorithm \cite{kuhn1955hungarian,munkres1957algorithms,burkard2012assignment,wolsey2020integer}. Therefore, for the constructed target set $S_3$, the resulting bijection $\phi$ minimizes the total Hamming distance
$\sum_{x\in S_2}D_H(x,\phi(x))$.

\cref{theorem:Mapping} establishes the optimal assignment for the constructed target set $S_3$. The mode-based selection of the fixed pattern $\tilde{a}$ serves as a heuristic for constructing $S_3$ by maximizing the overlap $|S_2\cap S_3|$, thereby reducing the number of basis states requiring permutation and the size of the subsequent assignment problem.
	
A further (computationally more expensive) extension is to additionally search over the choice of the fixed pattern $\tilde{a}$. This leads to the discrete optimization problem
\begin{equation}
	\min_{\tilde{a}} \; \Bigg( \min_{\phi: S_2 \to S_3(\tilde{a})} \sum_{x\in S_2} D_H(x,\phi(x)) \Bigg),
\end{equation}
where for each fixed $\tilde{a}$ the inner problem is a linear assignment problem.

Since $|S_3(\tilde{a})| = 2^n$ for all admissible $\tilde{a}$, this induces a finite family of assignment problems indexed by $\tilde{a}$. In this work, we adopt a mode-based heuristic to select $\tilde{a}$, followed by solving the corresponding optimal assignment via the Hungarian algorithm. If multiple optimal bijections exist, a canonical choice may be obtained via deterministic	tie-breaking rules or by a small perturbation $C_{jk}\mapsto C_{jk}+\epsilon r_{jk}$ with $\epsilon>0$ arbitrarily small and distinct perturbation $r_{jk}$, which preserves optimality while ensuring uniqueness of the minimizer \cite{burkard2012assignment,schrijver2003combinatorial,korte2008combinatorial}.

\section{Coherent Permutation Using Multi-Controlled X Gates}\label{sec:Permutation}

In this section, we introduce a coherent permutation of amplitudes among basis states using MCX gates. Here, \textit{coherent} refers to a unitary (reversible) transformation that preserves superposition and relative phases, without measurement or state collapse.

Consider a $P$-qubit quantum state $|\psi\rangle = \sum_{i=0}^{2^P-1} \psi_i |i\rangle$, where the computational basis is indexed by binary strings over the index set $G = \{0, \dots, P-1\}$ (see \cref{eq:Binary_rep,fig:1_Qubit_ordering}). Any basis state $|a\rangle$ is represented by a binary string $a = (a_i)_{i \in G}$. We begin by defining a primitive operation that swaps amplitudes between two basis states differing in a single qubit.
\begin{definition}\label{definition:Swapping}
	\textnormal{A\_SWAP}$(a, b)$:  
	Let $a, b \in \{0,1\}^P$ be binary strings such that they differ only at qubit $t \in G$, i.e.,
	\[
	a_i = b_i \;\; \forall i \in G \setminus \{t\}, \quad a_t \oplus b_t = 1.
	\]
	Define the common control string $c = (c_i)_{i \in G \setminus \{t\}}$ where $c_i = a_i = b_i$. Then the amplitudes corresponding to $|a\rangle$ and $|b\rangle$ are swapped via
	\[
	\psi_a |a\rangle + \psi_b |b\rangle 
	\;\longmapsto\;
	\psi_b |a\rangle + \psi_a |b\rangle,
	\]
	by applying the gate
	\[
	C^{|c\rangle}_{G \setminus \{t\}} X_{|t\rangle}.
	\]
\end{definition}
Note that A\_SWAP permutes amplitudes without affecting amplitudes in other basis states.

\begin{definition}[Coherent permutation]\label{definition:Permutation}
	Let $S_2,S_3\subset S_1 = \{0,1\}^P$ with $|S_2|=|S_3|$, and let
	$\phi:S_2\rightarrow S_3$ be a bijection satisfying
	\[
	\phi(x) = x, \qquad x \in S_2 \cap S_3.
	\]
	Define
	\[
	S_2' = S_2\setminus S_3, \qquad	S_3'=S_3\setminus S_2.
	\]
	Then $\phi$ restricts to a bijection $\phi: S_2' \rightarrow S_3'$.	For each $a\in S_2'$, choose a shortest Hamming path
	\[
	\mathcal P_a:\quad
	a=a^{(0)}\rightarrow\cdots\rightarrow a^{(m_a)}=\phi(a),
	\]
	where $m_a=D_H(a,\phi(a)),$ such that
	\[
	D_H(a^{(k)},a^{(k+1)})=1,
	\]
	and
	\[
	a^{(k)}\notin S_2\cap S_3, \qquad k=0,\ldots,m_a-1.
	\]
	
	The corresponding local walk operator is defined by
	\[
	\mathcal W_a = \prod_{k=0}^{m_a-1} \mathrm{A\_SWAP} \!\left(a^{(k)},a^{(k+1)}\right),
	\]
	where the product is ordered along the path.
	Each $\mathcal W_a$ implements a cyclic permutation of the basis states
	along the ordered vertex set
	\[
	(a^{(0)},a^{(1)},\ldots,a^{(m_a)}),
	\]
	transporting the amplitude initially at $a^{(0)}$ to $a^{(m_a)} = \phi(a)$. 
	
	The coherent permutation operator is
	\[
	\mathrm{A\_PERMUTE} = \prod_{a\in S_2'}	\mathcal W_a,
	\]
	where the ordering of the local walk operators is specified by an admissible
	routing schedule (Definition~\ref{definition:RoutingSchedule}).
	
\end{definition}
	
\begin{definition}[Admissible routing schedule]
	\label{definition:RoutingSchedule}
	
	Let
	\[
	\mathcal W_{a_1},\mathcal W_{a_2},\ldots,\mathcal W_{a_r},
	\qquad
	r=|S_2'|,
	\]
	be an ordering of the local walk operators. The ordering is called
	\emph{admissible} if it satisfies the following:
	
	\begin{enumerate}
		
		\item Each local walk operator is executed completely before the next local walk begins.
		
		\item A source basis state in $S_2'$ may be reused as an intermediate vertex only after its assigned amplitude has reached its destination.
		
		\item A target basis state in $S_3'$ may be used as an intermediate vertex only before its assigned amplitude arrives.
		
		\item Intermediate basis states outside $S_2 \cup S_3$ may be reused arbitrarily.
		
		\item Distinct routing paths may intersect and share intermediate basis states. No path-disjointness condition is required, provided Conditions (1)--(4) are satisfied.
		
	\end{enumerate}	
\end{definition}

\begin{theorem}[Global permutation induced by coherent routing]
	\label{theorem:A_PERMUTE}
	Under the coherent permutation construction of \cref{definition:Permutation}, suppose that the selected Hamming paths admit an admissible routing schedule according to \cref{definition:RoutingSchedule}. Then $\mathrm{A\_PERMUTE}$ is a permutation unitary inducing a permutation $\sigma: S_1 \mapsto S_1$ such that
	\[
	\sigma(a)=\phi(a),\quad \forall a\in S_2.
	\]
	Consequently,
	\[
	\sigma(S_2)=S_3, \quad	\sigma(S_1\setminus S_2)=S_1\setminus S_3.
	\]
	In general, the restriction of $\sigma$ to $S_1 \backslash S_2$ is a nontrivial complementary permutation induced by the same $\mathrm{A\_SWAP}$ network.
\end{theorem}

\textit{Proof:}
Each $\mathrm{A\_SWAP}$ is a transposition of computational basis states, and hence $\mathrm{A\_PERMUTE}$, being a finite product of such operations, is a permutation unitary. Therefore, there exists a unique permutation $\sigma: S_1 \mapsto S_1$ such that
\[
	\mathrm{A\_PERMUTE}|x\rangle=|\sigma(x)\rangle, \qquad \forall x\in S_1.
\]
It remains to show that $\sigma|_{S_2} = \phi$.

Let $(a_1, \dots, a_r)$ be the ordering of the elements of $S_2'$ specified by the admissible routing schedule, where $r = |S_2'|$ and $\{a_1, \dots, a_r\} = S_2'$. Then define \begin{equation}
U_0=I, \; U_q=\mathcal W_{a_q}\cdots\mathcal W_{a_1}, \; q=1,\ldots,r.
\end{equation}
We prove by induction on $q$ that
\begin{equation}\label{eq:th_4_proof_1}
U_q|a_\ell\rangle=|\phi(a_\ell)\rangle, \qquad \ell=1,\ldots,q,
\end{equation}
and
\begin{equation}\label{eq:th_4_proof_2}
U_q|x\rangle=|x\rangle, \qquad x\in S_2\cap S_3.
\end{equation}
For $q=0$, both statements are immediate. 
Assume \cref{eq:th_4_proof_1,eq:th_4_proof_2} hold for $q-1$. Since $a_q$ is an unprocessed source, Condition 2 of \cref{definition:RoutingSchedule} prevents $a_q$ from being used as an intermediate vertex in any of the preceding local walks. Hence
\[
U_{q-1}|a_q\rangle = |a_q\rangle.
\]
By the definition of the local walk,
\begin{equation}\label{eq:th_4_proof_3}
	\mathcal{W}_{a_q} |a_q\rangle = |\phi(a_q)\rangle.
\end{equation}
and therefore
\begin{equation}\label{eq:th_4_proof_4}
	U_{q}|a_q\rangle = \mathcal{W}_{a_q} U_{q-1} |a_q\rangle = | \phi(a_q)\rangle.
\end{equation}
For every $\ell < q$, the induction hypothesis gives
\[
U_{q-1}|a_\ell\rangle = |\phi(a_{\ell})\rangle.
\]
Since the assigned target $\phi(a_\ell)$ has already been reached, admissibility prevents it from being used in any subsequent local walk. Thus
\[
\mathcal{W}_{a_q}|\phi(a_\ell)\rangle = |\phi(a_\ell)\rangle, \qquad \ell < q,
\]
and hence
\begin{equation}\label{eq:th_4_proof_5}
	U_q |a_\ell \rangle = \mathcal{W}_{a_q} U_{q-1}|a_\ell\rangle = |\phi(a_\ell)\rangle, \qquad \ell < q.
\end{equation}
Finally, no state in $S_2 \cap S_3$ occurs on any routing path. Therefore
\[
\mathcal{W}_{a_q}|x\rangle = |x\rangle, \qquad x \in S_2 \cap S_3,
\]
and, using the induction hypothesis,
\begin{equation}\label{eq:th_4_proof_6}
	U_q|x\rangle = \mathcal{W}_{a_q} U_{q-1}|x \rangle = |x \rangle.
\end{equation}
\cref{eq:th_4_proof_4,eq:th_4_proof_5,eq:th_4_proof_6} establish \cref{eq:th_4_proof_1,eq:th_4_proof_2} for $q$, completing the induction.

Setting $q=r$, we obtain
\begin{equation}\label{eq:a_in_S_2'}
\sigma(a)=\phi(a), \qquad a\in S_2',
\end{equation}
and
\begin{equation}\label{eq:x_in_S_2_and_S_3}
\sigma(x)=x, \qquad x\in S_2\cap S_3.
\end{equation}
Since
\[
S_2=S_2'\cup(S_2\cap S_3)
\]
and $\phi(x)=x$ on $S_2\cap S_3$, it follows from \cref{eq:a_in_S_2',eq:x_in_S_2_and_S_3} that
\[
\sigma(a)=\phi(a), \quad a\in S_2.
\]
Therefore,
\[
\sigma(S_2)=\phi(S_2)=S_3.
\]
Since $\sigma$ is a permutation of $S_1$,
\[
\sigma(S_1 \backslash S_2) = S_1 \backslash S_3.
\]
The action on $S_1 \backslash S_2$ is therefore the complementary permutation induced by the same $\mathrm{A\_SWAP}$ network and is not, in general, the identity.

For the proposed construction, the selected shortest Hamming paths are required to admit an admissible routing schedule. If no admissible schedule can be identified, the assignment $\phi$, the target set $S_3$, or the selected shortest paths may be reconsidered until an admissible routing configuration is obtained. Accordingly, \cref{theorem:A_PERMUTE} guarantees the correctness of the resulting coherent permutation whenever such a configuration exists.

We now consider a small example to illustrate the coherent permutation. 
\begin{example}\label{example:coherent_permutation}
	Consider three qubits as $|j_2 j_1 j_0\rangle$. Let the bijection 
	\[
	\begin{split}
		\phi: \; &S_2 = \{000, 001, 100, 111\} \mapsto \\
		&S_3 = \{000, 001, 010, 011\},
	\end{split}
	\]
	where $S_2' = \{100, 111\}$, $S_3' = \{010, 011\}$. We can choose the following shortest Hammings paths for the elements of $S_2'$:
	\[
	\mathcal{P}_{100} = 100 \mapsto 110 \mapsto 010, \quad \mathcal{P}_{111} = 111 \mapsto 011.
	\]
	These paths form an admissible routing schedule according to \cref{definition:RoutingSchedule}, since the common states $000$ and $001$ are not used as intermediate vertices and the two routing operations do not interfere with previously assigned destinations. The corresponding \textnormal{A\_PERMUTE} operator (see \cref{theorem:A_PERMUTE}) is
	\begin{equation}\label{eq:example_permutation}
		\begin{aligned}
			&\textnormal{A\_PERMUTE} |x\rangle\\ 
			&= C_{|j_1j_0\rangle}^{|11\rangle}X_{|j_2\rangle} \times C_{|j_1j_0\rangle}^{|10\rangle}X_{|j_2\rangle} \times C_{|j_2j_0\rangle}^{|10\rangle}X_{|j_1\rangle} |x\rangle\\
			&= C_{|j_1\rangle}^{|1\rangle}X_{|j_2\rangle} \times C_{|j_2j_0\rangle}^{|10\rangle}X_{|j_1\rangle} |x\rangle
		\end{aligned}
	\end{equation}
\end{example}

Note that the two MCX gates in the equation \cref{eq:example_permutation} are compressed according to \cref{theorem:MCX_Composition}. Generalizing this example, we now consider the inverse construction: expressing a single MCX gate as a composition of MCX gates.

\begin{theorem}\label{theorem:MCX_Inverse}
	Let $|\tilde{a}\rangle$ be a control state defined on indices $F \subset G$, and $|G \setminus F| = n$. Then there exists a set of control states $S_2 = \{a^j\}_{j=0}^{2^n-1}$ such that
	\[
	a^j_i = \tilde{a}_i \quad \forall i \in F,
	\quad
	\{a^j_{G \setminus F}\}_{j=0}^{2^n-1} = \{0,1\}^n,
	\]
	and
	\[
	C_F^{|\tilde{a}\rangle} X_{|t\rangle} = \prod_{j=0}^{2^n-1} C_G^{|a^j\rangle} X_{|t\rangle}.
	\]
\end{theorem}
\textit{Proof:} Let	$H = G\setminus F, |H|=n.$ Since the free subsystem spans all binary strings of length \(n\), there exists an orthonormal basis $\{|a'^j\rangle\}_{j=0}^{2^n-1} = \{|0,1\rangle^{\otimes n}\},$ satisfying
	\[
	\sum_{j=0}^{2^n-1} |a'^j\rangle\langle a'^j| =	I_H.
	\]
	Then
	\[
	\resizebox{\columnwidth}{!}{$
	C_F^{|\tilde a\rangle}X_{|t\rangle}	= |\tilde a\rangle\langle\tilde a| \otimes I_H \otimes X_{|t\rangle} + \Bigl(I- |\tilde a\rangle\langle\tilde a|\Bigr)	\otimes I_H	\otimes I_{|t\rangle}.
	$}
	\]
	Substituting $I_H	= \sum_{j=0}^{2^n-1} |a'^j\rangle\langle a'^j|,$ gives
	\[
	\begin{aligned}
		C_F^{|\tilde a\rangle}X_{|t\rangle} = \sum_{j=0}^{2^n-1} \Big(
		& |\tilde a\rangle\langle\tilde a| \otimes |a'^j\rangle\langle a'^j| \otimes X_{|t\rangle}\\
		+ &	(I-|\tilde a\rangle\langle\tilde a|) \otimes |a'^j\rangle\langle a'^j| \otimes I_{|t\rangle}\Big).
	\end{aligned}
	\]
	Now define $|a^j\rangle	= |\tilde a\rangle\otimes |a'^j\rangle.$ By construction, $a_i^j=\tilde a_i, i\in F,$	and $\{a^j_{G\setminus F}\}_{j=0}^{2^n-1} = \{0,1\}^n.$ Using the orthogonality relation $\langle a'^j|a'^k\rangle = \delta_{jk},$ the summation above is precisely the product
	\[
	\begin{split}
		C_F^{|\tilde a\rangle}X_{|t\rangle} = \prod\limits_{j=0}^{2^n-1} \! &\left[ |\tilde{a}\rangle \langle \tilde{a} | \! \otimes \! |a'^j\rangle \langle a'^j| \! \otimes \! X_{|t\rangle} \! + \right. \\
		&\left. (I \! -\! (|\tilde{a}\rangle \langle \tilde{a} | \! \otimes \! |a'^j\rangle \langle a'^j|)) \! \otimes \! I_{|t\rangle} \right].
	\end{split}
	\]
	Hence
	\[
	\begin{split}
		C_F^{|\tilde a\rangle}X_{|t\rangle} &= \prod\limits_{j=0}^{2^n-1} \left[ |a^j\rangle\langle a^j| \otimes X_{|t\rangle} + \right.\\
		&\hspace{2cm}\left.(I - |a^j\rangle\langle a^j|) \otimes I_{|t\rangle} \right],\\
		&= \prod_{j=0}^{2^n-1} C_G^{|a^j\rangle}X_{|t\rangle}.
	\end{split}
	\]
	
Operationally, this implies that a single MCX gate can be interpreted as simultaneously performing A\_SWAP operations across all pairs of basis states whose control patterns belong to $S_2$. Note that this \cref{theorem:MCX_Inverse} becomes highly relevant in further reducing the implementation cost of A\_PERMUTE operator as discussed in \cref{sec:resource_accounting}. To demonstrate this process, consider an example of three qubits $(j_2, j_1, j_0)$ and a quantum state vector $[\psi_i]_{i=0}^7$. The CONTROLLED-NOT gate
\[
C^{|1\rangle}_{|j_1\rangle} X_{|j_0\rangle} = C^{|01\rangle}_{|j_2 j_1\rangle} X_{|j_0\rangle}
\;\times\;
C^{|11\rangle}_{|j_2 j_1\rangle} X_{|j_0\rangle},
\]
permutes the amplitudes as
\[
\begin{bmatrix}
	\psi_0\\
	\psi_1\\
	\psi_2\\
	\psi_3\\
	\psi_4\\
	\psi_5\\
	\psi_6\\
	\psi_7
\end{bmatrix}
\;\longmapsto\;
\begin{bmatrix}
	\psi_0\\
	\psi_1\\
	\psi_3\\
	\psi_2\\
	\psi_4\\
	\psi_5\\
	\psi_7\\
	\psi_6
\end{bmatrix}.
\]

\begin{figure*}[t]
	\centering
	\begin{subfigure}[t]{0.23\textwidth}
		%\centering
		\includegraphics[width=\textwidth]{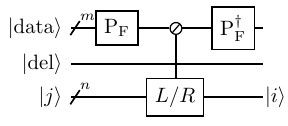}
		\caption{} 
		\label{fig:5a_Reduction}
	\end{subfigure}
	\hfill
	\begin{subfigure}[t]{0.37\textwidth}
		%\centering
		\includegraphics[width=\textwidth]{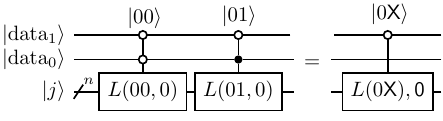}
		\caption{}
		\label{fig:5b_Reduction}
	\end{subfigure}
	\hfill
	\begin{subfigure}[t]{0.38\textwidth}
		%\centering
		\includegraphics[width=\textwidth]{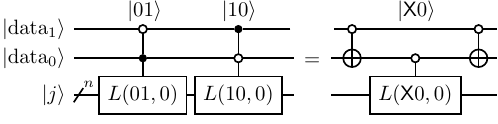}
		\caption{}
		\label{fig:5c_Reduction}
	\end{subfigure}
	\\
	\begin{subfigure}[t]{0.28\textwidth}
		%\centering
		\includegraphics[width=\textwidth]{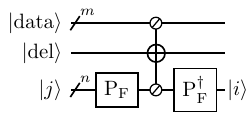}
		\caption{}
		\label{fig:5d_Reduction}
	\end{subfigure}
	\hfill
	\begin{subfigure}[t]{0.4\textwidth}
		%\centering
		\includegraphics[width=\textwidth]{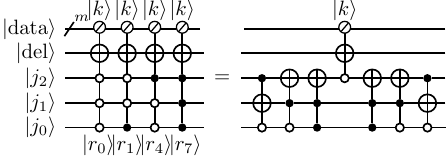}
		\caption{}
		\label{fig:5e_Reduction}
	\end{subfigure}
	\hfill
	\begin{subfigure}[t]{0.3\textwidth}
		%\centering
		\includegraphics[width=\textwidth]{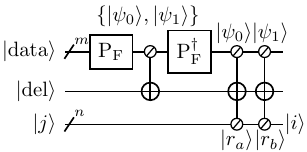}
		\caption{}
		\label{fig:5f_Reduction}
	\end{subfigure}
	\caption{\subref{fig:5a_Reduction} Circuit representation of combined shift operation with permutation of amplitudes $\text{A\_PERMUTE}_\text{F}$ (represented by $\text{P}_{\text{F}}$). \subref{fig:5b_Reduction} Circuit for \cref{example_1}, showing the combined left shift of data elements $\{\psi_0, \psi_1\}$ in basis states $\{|00\rangle, |01\rangle\}$ by one column. \subref{fig:5c_Reduction} Circuit for \cref{example_2}, illustrating permutation of amplitudes and left shift of data elements $\{\psi_1, \psi_2\}$ by one column. \subref{fig:5d_Reduction} Circuit representation of combined deletion with permutation of rows $\text{P}_{\text{F}}$. \subref{fig:5e_Reduction} Circuit for \cref{example_4}, showing deletion of data element $|k\rangle$ in rows $\{|r_0\rangle, |r_1\rangle, |r_4\rangle, |r_7\rangle\}$. Note that the MCX gates here within $\text{P}_{\text{F}}$ can also be compressed as in \cref{eq:example_permutation}, but retained to show the full walk operator \cref{example:coherent_permutation}. \subref{fig:5f_Reduction} Circuit for \cref{example_5}, illustrating the insertion of data elements $\{|\psi_0\rangle, |\psi_1\rangle\}$ in rows $|r_a\rangle, |r_b\rangle$, respectively, along with $\text{P}_{\text{F}}$.}
	\label{fig:5_Reduction}
\end{figure*}

\section{Optimized Index Mapping Oracle}\label{sec:Optimized_Index_Mapping_Oracle}
We have seen that index mapping oracles $O_{\text{shift}}$ and $O_{\text{del}}$ \cref{sec:Index_mapping_oracle} can shift and delete the data elements in the block encoded matrix. We can compress the composition of $2^n$ MCX gates when the control set $S_2$ exhibits a fixed set of indices $F \subset G$ and satisfies the structural constraints in \cref{sec:MCX_Composition}. Otherwise, we can determine an arbitrary $F$ and find a set $S_3$ such that $\phi: S_2 \mapsto S_3$ \cref{sec:Mapping}. The choice of fixed indices $F$ determines the qubits on which the MCX gates are applied. Therefore, we can use this as an advantage in superconducting quantum hardware and apply the MCX gates on nearest-neighbor qubits reducing the control complexity. 

In this section, we illustrate the optimized index-mapping operations through several examples arising in sparse-matrix block encoding, covering a range of representative applications. We further present hardware-level resource analyses for the presented examples to assess the practical impact of the proposed constructions. The analysis is performed for IBM superconducting quantum processor with heavy-hex and square lattice topologies. The circuits are transpiled to IBM superconducting hardware topologies and evaluated using two-qubit depth as the primary resource metric. Two-qubit depth is defined as the depth of native two-qubit operations in the transpiled circuit after accounting for device connectivity constraints, and corresponds to the critical path of sequential two-qubit gates, which typically provides the dominant contribution to runtime and error accumulation on superconducting hardware.

\subsection{Shift}
In the combined operation of shift oracles \cref{sec:Index_mapping_oracle}, the objective is to get nearest-neighbor MCX gates. To achieve this, the choice of $F$ comes from choosing the qubits in $|\text{data}\rangle$ closer to matrix qubits $|j\rangle$. The choice of $F$ determines the mapping $\phi: S_2 \mapsto S_3$. Then the amplitudes are permuted using $\text{A\_PERMUTE}_\text{F}$ (refer \cref{definition:Permutation}). The visualization of this combined shift operation is shown in \cref{fig:5a_Reduction}. After shifting, the amplitudes are rearranged back to retain the original order, avoiding confusion in subsequent operations.

We now present some examples for intuitive understanding. Consider block encoding an $8 \times 8$ matrix with three matrix qubits $(|j\rangle)$ and two data qubits $(|\text{data}_1\rangle, |\text{data}_0\rangle)$. Let $v_{\text{data}} = [\psi_0, \psi_1, \psi_2, \psi_3]^T$, basis states for $|\text{data}\rangle$ are $(00, 01, 10, 11)$. 
\begin{example}\label{example_1}
	The data elements $\{\psi_0, \psi_1\}$ are to be shifted left by one column using the operators $L(01, 0)L(00, 0)$, corresponding to a many-to-one mapping.
\end{example}
\textit{Solution}: The control set $S_2 = \{00, 01\}$ satisfies the structural constraints in \cref{theorem:MCX_Composition} such that $|S_2| = 2$, fixed index $F = \{|\text{data}_1\rangle\}$ for first composition of MCX gates as in \cref{eq:MCX_composition,fig:4_Commute}. Then the combined shift operation is given by 
\begin{equation}
	L(01, 0)L(00, 0) = L(0\mathsf{X}, 0),
\end{equation}
where $\mathsf{X}$ denote no control gate on qubit $|\text{data}_0\rangle$. The corresponding circuit representation is shown in \cref{fig:5b_Reduction}.  

\begin{example}\label{example_2}
	The data elements $\{\psi_1, \psi_2\}$ are to be shifted left by one column using the operators $L(01, 0)L(10, 0)$, corresponding to a many-to-one mapping.
\end{example}
\textit{Solution}: The control states $S_2 = \{01, 10\}$ does not satisfy the constraints in \cref{theorem:MCX_Composition}. So we choose the fixed index $F = \{|\text{data}_0\rangle\}$ for first composition of MCX gates as in \cref{eq:MCX_composition,fig:4_Commute}. We apply the $\text{A\_PERMUTE}_\text{F}$ (see \cref{fig:5a_Reduction}), where the amplitudes are swapped $\psi_0 |00\rangle + \psi_1|01\rangle \mapsto \psi_1 |00\rangle + \psi_0|01\rangle$ using a CNOT (control value is 0) gate, as shown in \cref{fig:5c_Reduction}. After this permutation, the combined shift operator can be applied as:
\begin{equation}
	L(00, 0)L(10, 0) = L(\mathsf{X}0, 0).
\end{equation}

For the case of $n=3$ matrix qubits, we performed hardware resource analysis of the circuit (see \cref{fig:5c_Reduction}) before (with only shift oracles) and after (along with coherent permutation and compression) optimization of the index-mapping oracle. As shown in \cref{table:resource_analysis_example_2}, the optimized construction achieves a significant reduction in two-qubit depth for both heavy-hex and square-lattice topologies.
\begin{table}[ht]
	\centering
	\resizebox{\columnwidth}{!}{$
	\begin{tabular}{|c|c|}
		\hline
		\multicolumn{2}{|c|}{Hardware resource analysis of \cref{example_2}}\\
		\hline
		Before optimization & Hex $= 173$, Square $=144$\\
		\hline
		After optimization & Hex $= 36$, Square $=35$\\
		\hline
	\end{tabular}
	$}
	\caption{Hardware resource analysis of \cref{example_2} measured in terms of two-qubit depth. Here before optimization denote using only individual shift oracles, and after optimization denote coherent permutation and compressed shift oracle. Here `Hex' denotes the heavy-hex lattice topology of the Heron r3 processor (IBM Boston), and `Square' denotes the square lattice topology of the Nighthawk r1 processor (IBM Miami).}
	\label{table:resource_analysis_example_2}
\end{table}

\begin{example}\label{example_3}
	Assume a data vector $v_{\textnormal{data}}$ of 7 values padded with $0$: $v_{\textnormal{data}} = [\psi_0, \psi_1, \psi_2, \psi_3, \psi_4, \psi_5, \psi_6, 0]^T$. The task is to shift the data elements $\{\psi_0, \psi_5, \psi_6\}$ left by one column.
\end{example}
\textit{Solution}: According to \cref{theorem:MCX_Composition}, the control set $S_2$ should be a power of two $(2^n)$. Since the task involves three data elements, a naive approach would be to apply shift operation individually. Alternatively, one can exploit the $0$ in the data vector and perform a combined shift on four data elements $\{\psi_0, \psi_5, \psi_6, \psi_7\}$, as shifting $0$ does not affect the block-encoded matrix \cref{eq:diagonal_block_encoded}. The combined operation for these basis states requires permutation, as illustrated in \cref{fig:5a_Reduction}. 

Note that shifting a single data element left and then right results in the identity operation, i.e., $L\times R = I$ (refer \cref{fig:3a_Shifting,fig:3b_Shifting}). This property can be exploited when applying combined operations to simplify MCX gates.

\subsection{Delete}
We have seen how a single data element can be deleted in a specific row \cref{fig:3d_Deletion}. In this section, we generalize this to deletion in multiple rows. Consecutive delete operations of a single data element across multiple rows result in a composition of MCX gates with control and target on the same qubits. When the control states of such a composition satisfy the constraints in \cref{theorem:MCX_Composition}, they can be combined into a single MCX gate. Otherwise, one can choose a set of fixed indices $F \subset G$ and determine $\phi : S_2 \mapsto S_3$ (refer \cref{sec:Mapping}). The choice of $F$ comes from choosing the qubits in $|j\rangle$ close to $|\text{data}\rangle$ and $|\text{del}\rangle$ to achieve nearest-neighbor connectivity. The combined delete operation along with the permutation operator $\text{A\_PERMUTE}_\text{F}$ is presented in \cref{fig:5d_Reduction}.

\begin{example}\label{example_4}
	Consider three matrix qubits $|j_2j_1j_0\rangle$. The task is that the $k^{\text{th}}$ data element is to be deleted in rows $\{0, 1, 4, 7\}$ using the operators $D^{|k\rangle}_{|0\rangle} D^{|k\rangle}_{|1\rangle} D^{|k\rangle}_{|4\rangle} D^{|k\rangle}_{|7\rangle}$, corresponding to a one-to-many mapping.
\end{example}
\textit{Solution}: The set of control states $S_2 = \{000, 001, 100, 111\}$ does not have $F$ and hence does not satisfy the constraints in \cref{theorem:MCX_Composition}. We choose $F = \{|j_2\rangle\}$ and obtain $\phi: S_2 = \{0, 1, 4, 7\} \mapsto S_3 = \{0, 1, 2, 3\}$ (see \cref{sec:Mapping}) using the linear sum assignment algorithm \cite{crouse2016implementing}. The corresponding permutation of amplitudes (see \cref{example:coherent_permutation}) is given as
\[
\begin{aligned}
	000&\rightarrow000&\rightarrow000\\
	001&\rightarrow001&\rightarrow001\\
	100&\rightarrow C_{|j_1j_0\rangle}^{|10\rangle}X_{|j_2\rangle} \cdot C_{|j_2j_0\rangle}^{|10\rangle}X_{|j_1\rangle}&\rightarrow010\\
	111&\rightarrow C_{|j_1j_0\rangle}^{|11\rangle}X_{|j_2\rangle}&\rightarrow011
\end{aligned}
\]
The combined deletion is then given by 
\begin{equation}
	D^{|k\rangle}_{\{|0\rangle, |1\rangle, |4\rangle, |7\rangle\}} = C_{|\text{data}\rangle}^{|k\rangle}C^{|0\rangle}_{|j_2\rangle}X_{|\text{del}\rangle},
\end{equation}
After the deletion, the rows are permuted back to their original order. The circuit for this example is shown in \cref{fig:5e_Reduction}. 

For $|k\rangle = |10\rangle$, we performed the hardware resource analysis for the circuit (see \cref{fig:5e_Reduction}) reported in \cref{table:resource_analysis_example_4}. The results indicate once again that optimized construction achieves a significant reduction in two-qubit depth for both heavy-hex and square lattice topologies. 

\begin{table}[ht]
	\centering
	\resizebox{\columnwidth}{!}{$
	\begin{tabular}{|c|c|}
		\hline
		\multicolumn{2}{|c|}{Hardware resource analysis of \cref{example_4}}\\
		\hline
		Before optimization & Hex $= 704$, Square $=515$\\
		\hline
		After optimization & Hex $= 47$, Square $=47$\\
		\hline
	\end{tabular}
	$}
	\caption{Hardware resource analysis of \cref{example_4} measured in terms of two-qubit depth. Here before optimization denotes using only delete oracles, and after optimization denote using coherent permutation and compressed delete oracle. Here `Hex' denotes the heavy-hex lattice topology of the Heron r3 processor (IBM Boston), and `Square' denotes the square lattice topology of the Nighthawk r1 processor (IBM Miami).}
	\label{table:resource_analysis_example_4}
\end{table}
\begin{algorithm*}[t]
	\caption{Block encoding of sparse matrix via coherent permutation}
	\label{alg:BlockEncoding}
	
	\KwIn{Sparse matrix $A\in\mathbb C^{N\times N}$}
	
	\KwOut{Block-encoding unitary $U_A$ satisfying $\alpha \langle 0|U_A|0\rangle=A$}
	
	\BlankLine
	
	\textbf{Data preprocessing}
	\begin{enumerate}[itemsep=-0.5pt, topsep=1pt]
		\item Extract the nonzero entries $S_{d}$ along each diagonal offset $d\in\mathrm{Diag}$ (see \cref{eq:unique_set}).
		\item Construct $v_{\mathrm{data}}$, $v_{\mathrm{sign}}$ (\cref{eq:v_data,eq:v_sign}).
		\item Associate to each data element $(v_{\mathrm{data},k}, v_{\mathrm{sign},k}, d_k, S_r(k), |k\rangle_{\mathrm{data}})$ (see \cref{table:data_structure}).
	\end{enumerate}
	
	\BlankLine
	
	\textbf{State preparation}
	\begin{enumerate}[itemsep=-0.5pt, topsep=1pt]
		\item Construct PREP and UNPREP oracles (\cref{sec:State_prep_oracle}).
	\end{enumerate}
	
	\BlankLine
	
	\textbf{Index mapping optimization}
	\begin{enumerate}[itemsep=-0.5pt, topsep=1pt]
		\item Determine the required shift, delete, and insert operators for each data element (\cref{sec:Index_mapping_oracle}).
		\item Identify compressible control patterns and apply MCX-compression (\cref{theorem:MCX_Composition}).
		\item If a compressed operator cannot be implemented directly, construct a target set $S_3$ from a source set $S_2$ using the mode-based heuristic (\cref{theorem:Mapping}).
		\item Solve the linear assignment problem (\cref{eq:Optimization}) to obtain the bijection $\phi:S_2\rightarrow S_3$.
		\item Determine admissible routing paths and synthesize the coherent permutation $\mathrm{A\_PERMUTE}$ (\cref{sec:Permutation}).
	\end{enumerate}
	
	\BlankLine
	
	\textbf{Circuit synthesis}
	\begin{enumerate}[itemsep=-0.5pt, topsep=1pt]
		\item Get the combined shift, delete, and insert operators with the help of permutation operators to obtain the optimized index mapping oracle (see \cref{sec:Optimized_Index_Mapping_Oracle}).
		\item Assemble the PREP/UNPREP with optimized index mapping oracles to get the complete block encoding circuit (see \cref{fig:6a_Complete_circuit}).
		\item Multiply by the normalization factor $\alpha$ to recover the original matrix $A$ (\cref{theorem_1}).
	\end{enumerate}
\end{algorithm*}

Furthermore, a many-to-many mapping is also possible, where more than one data element can be deleted in multiple rows, provided that the set of rows to be deleted is common for all data elements.
\begin{figure*}[t]
	\centering
	\begin{subfigure}[t]{0.9\textwidth}
		%\centering
		\includegraphics[width=\textwidth]{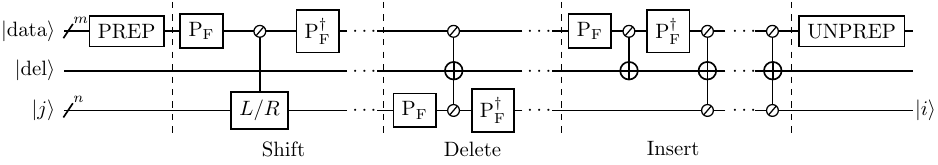}
		\caption{} 
		\label{fig:6a_Complete_circuit}
	\end{subfigure}
	\\
	\begin{subfigure}[t]{0.85\textwidth}
		%\centering
		\includegraphics[width=\textwidth]{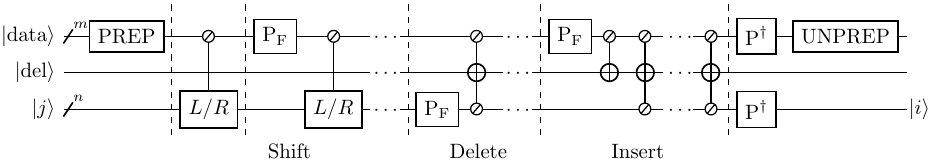}
		\caption{}
		\label{fig:6b_Complete_circuit}
	\end{subfigure}
	\caption{\subref{fig:6a_Complete_circuit} Circuit representation for block encoding with coherent permutation. The circuit is initialized with $m \; |\text{data}\rangle$, one $|\text{del}\rangle$ and $n$ matrix ($|j\rangle$) qubits. It includes the PREP and UNPREP operators from the state preparation oracle (see \cref{sec:State_prep_oracle}). The combined shift, delete, and insert operators \cref{fig:5a_Reduction,fig:5d_Reduction,fig:5f_Reduction} are incorporated along with the corresponding $\text{A\_PERMUTE}_\text{F}$ (represented as $\text{P}_{\text{F}}$) operators for permutation of amplitudes for a chosen $F$. The dots indicate repeating structures. \subref{fig:6b_Complete_circuit} A second circuit architecture allows the amplitude order to evolve throughout the computation, avoids intermediate inverse permutations and a single inverse permutation $\mathrm{P}^{\dagger}$ is applied at the end.}
	\label{fig:6_Complete_circuit}
\end{figure*}
\subsection{Insert}\label{sec:Insert}
In this section, we introduce an insert operator which places the data elements into one or few rows without requiring exponential MCX gates as illustrated in \cref{example_5}

\begin{example}\label{example_5}
	Consider an example of block encoding a matrix of $2^n \times 2^n$. The data elements $\{\psi_0, \psi_1\}$ are to be inserted in different rows $\{r_a, r_b\}$, respectively, corresponding to a one-to-one mapping.
\end{example}
\textit{Solution}: To insert a single data element $\psi_0$ in row $r_a$, we define 
\begin{equation}
		\resizebox{\columnwidth}{!}{$
I^{|\psi_0\rangle}_{|r_a\rangle} =  D^{|\psi_0\rangle}_{|r_a\rangle} \left(\prod\limits_{k=0}^{2^n-1} D^{|\psi_0\rangle}_{|k\rangle}\right) = D^{|\psi_0\rangle}_{|r_a\rangle} \left( C^{|\psi_0\rangle}_{|\text{data}\rangle}X_{|\text{del}\rangle} \right),
$}
\end{equation} 
where $\left(\prod\limits_{k=0}^{2^n-1} D^{|\psi_0\rangle}_{|k\rangle}\right)$ means deleting in all rows and it is compressed to single MCX gate as presented in \cref{eq:n=p,theorem:MCX_Composition}. Since the proposed block encoding method places each data element in every row \cref{eq:diagonal_block_encoded}, we first delete the data element from all rows and then apply deletion on the desired row $|r_a\rangle$. This effectively inverts the deletion on the desired row, resulting in the insertion of the data element in that row alone.

To insert a set of data elements $\{\psi_0, \psi_1\}$ into rows $\{|r_a\rangle, |r_b\rangle\}$, respectively, we formulate 
\begin{equation}
			\resizebox{\columnwidth}{!}{$
	\begin{split}
		I^{|\psi_0\rangle}_{|r_a\rangle}I^{|\psi_1\rangle}_{|r_b\rangle} &=  D^{|\psi_0\rangle}_{|r_a\rangle} \left(\prod\limits_{k=0}^{2^n-1} D^{|\psi_0\rangle}_{|k\rangle}\right)  D^{|\psi_1\rangle}_{|r_b\rangle} \left(\prod\limits_{k=0}^{2^n-1} D^{|\psi_1\rangle}_{|k\rangle}\right)\\
		&=  D^{|\psi_0\rangle}_{|r_a\rangle} D^{|\psi_1\rangle}_{|r_b\rangle} \left(\prod\limits_{k=0}^{2^n-1} D^{\{|\psi_0\rangle, |\psi_1\rangle\}}_{|k\rangle}\right).
	\end{split}
	$}
\end{equation}
The circuit representation for this task is shown in \cref{fig:5f_Reduction}, where $\text{A\_PERMUTE}_\text{F}$ (represented as $\text{P}_{\text{F}}$) is included (if needed) for permutation of amplitudes for a chosen $F$.

For the instance $|\psi_0\rangle = |01\rangle, |\psi_1\rangle = |10\rangle, n = 3, |r_a\rangle = |101\rangle, |r_b\rangle = |011\rangle$, we performed the hardware resource analysis of the circuit (\cref{fig:5f_Reduction}) comparing the original implementation, consisting only of delete oracles, with the optimized implementation incorporating insert and coherent permutation operators. The results, reported in \cref{table:resource_analysis_example_5}, show that for this instance the introduction of insert and coherent permutation operations leads to a significant reduction in the transpiled two-qubit depth.

\begin{table}[ht]
	\centering
	\resizebox{\columnwidth}{!}{$
		\begin{tabular}{|c|c|}
			\hline
			\multicolumn{2}{|c|}{Hardware resource analysis of \cref{example_5}}\\
			\hline
			Before optimization & Hex $= 2234$, Square $=2069$\\
			\hline
			After optimization & Hex $= 330$, Square $=272$\\
			\hline
		\end{tabular}
		$}
	\caption{Hardware resource analysis of \cref{example_5} measured in terms of two-qubit depth. Here before optimization denote using only delete oracles, and after optimization denote using insert and coherent permutation operators. Here `Hex' denotes the heavy-hex lattice topology of the Heron r3 processor (IBM Boston), and `Square' denotes the square lattice topology of the Nighthawk r1 processor (IBM Miami).}
	\label{table:resource_analysis_example_5}
\end{table}

\section{Complete Circuit}\label{sec:Complete_circuit}

In this section, we present the circuit architectures and resource accounting for complete block encoding of sparse matrices. The workflow is summarized in \cref{alg:BlockEncoding}.

\subsection{Circuit Architectures}\label{sec:circuit_architectures}
A first architecture employing the optimized index mapping oracle (see \cref{sec:Optimized_Index_Mapping_Oracle}) is presented in \cref{fig:6a_Complete_circuit}. In this design, amplitudes are restored to a fixed ordering after every combined operation. Let the coherent permutation operator consist of $n$ MCX gates. Then,
\begin{equation}\label{eq:permute_order}
	\begin{split}
		\mathrm{A\_PERMUTE}
		&= \prod_{i=0}^{n-1}\mathrm{MCX}_i,\\
		\mathrm{A\_PERMUTE}^{\dagger} &= \prod_{i=0}^{n-1}\mathrm{MCX}_{n-1-i},
	\end{split}
\end{equation}
where the inverse permutation is implemented by applying the same MCX gates in reverse order.

A second architecture is illustrated in \cref{fig:6b_Complete_circuit}, where the state preparation oracle initializes amplitudes in an ordering already compatible with the first combined operation. In this case, intermediate inverse permutations are avoided; instead, the forward permutation effect is tracked implicitly through an evolving ordering of amplitudes, and a complete inverse permutation $\mathrm{A\_PERMUTE}^{\dagger}$ is applied at the end of the circuit. Whether this strategy yields a reduction in overall implementation cost compared with \cref{fig:6a_Complete_circuit} remains an open question.

\subsection{Resource Accounting}\label{sec:resource_accounting}
In this section, we discuss the cost comparison of full block encoding circuit. Let $\mathcal{C}(\cdot)$ denote an implementation cost metric, such as transpiled two-qubit depth, native two-qubit gate count, or a hardware-dependent weighted cost function.

For an original implementation consisting only of shift and delete operators,
\begin{equation}\label{eq:Cost_orig}
	\mathcal{C}_{\mathrm{orig}}
	=
	\mathcal{C}(\mathrm{PREP})
	+
	\mathcal{C}(\mathrm{UNPREP})
	+
	\sum_{r=1}^{R}
	\mathcal{C}(O_r),
\end{equation}
where $O_r\in \{O_{\mathrm{shift}}, O_{\mathrm{delete}}\}.$

The cost for architecture of \cref{fig:6a_Complete_circuit} is
\begin{equation}
	\begin{aligned}	\label{eq:Cost_opt_a}
		\mathcal{C}_{\mathrm{opt}}^{(a)} = & \; \mathcal{C}(\mathrm{PREP}) + \mathcal{C}(\mathrm{UNPREP})\\
		&+ \sum_{r=1}^{R_c} \Big(\mathcal{C}(P_r) + \mathcal{C}(O_r^{\mathrm{comp}}) + \mathcal{C}(P_r^\dagger)\Big),
	\end{aligned}
\end{equation}
where $R_c \le R$, $O_r^{\mathrm{comp}}$ denotes the $r^{\text{th}}$ compressed index-mapping operator obtained by combining shift, delete, and insert operations, and $P_r$ denotes the coherent permutation operator (\cref{theorem:A_PERMUTE}).

Similarly, the architecture of \cref{fig:6b_Complete_circuit} can be expressed as
\begin{equation}
	\label{eq:Cost_opt_b}
	\begin{split}
		\mathcal{C}_{\mathrm{opt}}^{(b)} =& \; \mathcal{C}(\mathrm{PREP}) + \mathcal{C}(\mathrm{UNPREP})\\
		& +	\sum_{r=1}^{R_c} \Big(\mathcal{C}(P_r) + \mathcal{C}(O_r^{\mathrm{comp}})\Big) + \mathcal{C}(P^\dagger),
	\end{split}
\end{equation}
where intermediate inverse permutations are avoided, rather a single inverse permutation $P^{\dag}$ is applied at the end.

The implementation cost of coherent permutation layers is generally nontrivial, since $\mathrm{A\_PERMUTE}$ itself may admit further compression through the same MCX-reduction techniques (see \cref{theorem:MCX_Inverse}) and coherent permutation introduced in this work. Consequently, the resource analysis depends on
\begin{enumerate}
	\item decomposition of MCX gates into native one- and two-qubit gates,
	\item ancilla requirements,
	\item routing overhead induced by hardware connectivity,
	\item possible further compression of coherent permutation networks.
\end{enumerate}

At the level of the complete block-encoding circuit, the relevant comparison is therefore between
\[
\mathcal{C}_{\mathrm{opt}}^{(a)},
\quad
\mathcal{C}_{\mathrm{opt}}^{(b)}
\quad\text{and}\quad
\mathcal{C}_{\mathrm{orig}}.
\]
For the representative examples considered in this work, hardware resource analyses (see \cref{table:resource_analysis_example_2,table:resource_analysis_example_4,table:resource_analysis_example_5}) indicate that the proposed optimization framework can lead to significant reductions in transpiled two-qubit depth for combined shift, delete, and insert operations, including connectivity-aware coherent permutation layers. More generally, the framework provides a systematic methodology for constructing and evaluating hardware-aware block-encoding circuits, where the overall implementation cost is determined by the interplay between MCX compression, permutation synthesis, and hardware connectivity constraints.

\section{Applications}\label{sec:Applications}
In this section, we present two examples of block encoding of sparse matrices: a complex tridiagonal matrix and a structured real matrix.

\subsection{Complex Tridiagonal Matrix}\label{sec:Tridiagonal_complex_matrix}
Consider a complex tridiagonal matrix $A \in \mathbb{C}^{2^n \times 2^n}$ of the form,
\begin{equation}\label{eq:complex_matrix}
	A = 
	\begin{bmatrix}
		z_2 & z_3\\
		z_1 & z_2 & z_3\\
		& \ddots & \ddots & \ddots\\
		&  & z_1 &z_2 & z_3\\
		& & &z_1 & z_2
	\end{bmatrix},
\end{equation}
where $z_1 = \psi_0 + \psi_1\textrm{i}, z_2 = \psi_2 + \psi_3\textrm{i}, z_3 = \psi_4+\psi_5\textrm{i}$ and $\psi_i \in \mathbb{R}$. The corresponding data vector is 
\[
v_{\text{data}} = [|\psi_0|, |\psi_1|, |\psi_2|, |\psi_3|, |\psi_4|, |\psi_5|]^T \; (\text{refer \cref{eq:v_data}})
\]
with sign vector
\[
\begin{split}
v_{\text{sign}} &= [\text{sgn}(\psi_0), \text{sgn}(\psi_1)\textrm{i}, \text{sgn}(\psi_2), \text{sgn}(\psi_3)\textrm{i}, \\
&\text{sgn}(\psi_4), \text{sgn}(\psi_5)\textrm{i}]^T \; (\text{refer \cref{eq:v_sign}}).
\end{split}
\]
State preparation requires $\left\lceil \log_2 6 \right\rceil = 3$ data qubits (refer \cref{eq:PREP}), where the state is padded with zeros. For block encoding, the data elements $\{\psi_0, \psi_1\}$ in basis states $\{|000\rangle, |001\rangle\}$ must be shifted left by one column and deleted in row $|r_0\rangle$ as in \cref{eq:L_Shift}. If required, amplitudes can be permuted to achieve nearest-neighbor MCX gate connectivity. Similarly, the data elements $\{\psi_4, \psi_5\}$ in basis states $\{|100\rangle, |101\rangle\}$ must be shifted right by one column and deleted in row $|r_{2^n-1}\rangle$ as in \cref{eq:R_Shift}. 

\begin{figure}[t]
	\centering
	\includegraphics[width=\columnwidth]{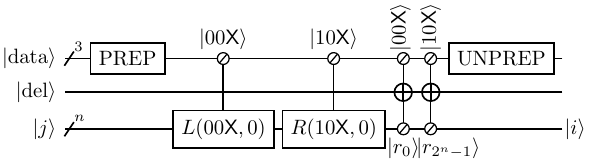}
	\caption{Circuit representation for block encoding complex tridiagonal matrix \cref{eq:complex_matrix}. Basis states $\{|000\rangle, |001\rangle\}$ compressed into $|00\mathsf{X}\rangle$ and $\{|100\rangle, |101\rangle\}$ compressed into $|10\mathsf{X}\rangle$ (refer \cref{theorem:MCX_Composition}).}
	\label{fig:7_Complex}
\end{figure}

The circuit representation of this construction is shown in \cref{fig:7_Complex} and provides a practical gate-level realization that can be directly employed within quantum algorithms. Note that zeros in the state vector can also be leveraged for combined shifting (see \cref{example_3}), thereby further reducing the control overhead of the MCX gates.

\begin{table}[ht]
	\centering
	\begin{tabular}{|l|}
		\hline
		\multicolumn{1}{|c|}{Block encoding operations}\\
		\hline
		$\psi_0, O_{\text{shift}}(0000, d=5), D^{|0000\rangle}_{\{0-9,30,31\}}$\\[2pt]
		$\psi_1, O_{\text{shift}}(0001, d=1), D^{|0001\rangle}_{\{0,5,10,15,20,25,30,31\}}$\\
		$\tilde{\psi}_2, D^{|0010\rangle}_{\{0-4\}}$\\
		$\tilde{\psi}_3$\\
		$\psi_4, O_{\text{shift}}(0100, d=-1), D^{|0100\rangle}_{\{4,9,14,19,24,29,30,31\}}$\\[2pt]
		$\psi_5, O_{\text{shift}}(0101, d=-5), D^{|0101\rangle}_{\{0-4,25-31\}}$\\[2pt]
		$\psi_6, O_{\text{shift}}(0110, d=6),I^{|0110\rangle}_{6}$\\[2pt]
		$\psi_7, O_{\text{shift}}(0111, d=9),I^{|0111\rangle}_{10}$\\[2pt]
		$\psi_8, O_{\text{shift}}(1000, d=11),I^{|1000\rangle}_{12}$\\[2pt]
		$\psi_9, O_{\text{shift}}(1001, d=14),I^{|1001\rangle}_{16}$\\[2pt]
		$\psi_{10}, O_{\text{shift}}(1010, d=16),I^{|1010\rangle}_{18}$\\[2pt]
		$\psi_{11}, O_{\text{shift}}(1011, d=19),I^{|1011\rangle}_{22}$\\[2pt]
		$\psi_{12}, O_{\text{shift}}(1100, d=21),I^{|1100\rangle}_{24}$\\[2pt]
		$\psi_{13}, O_{\text{shift}}(1101, d=24),I^{|1101\rangle}_{28}$\\
		$\psi_{14}=0$\\
		$\psi_{15}=0$\\
		\hline
	\end{tabular}
	\caption{Block encoding operations for \cref{fig:8a_Matrix}, with data vector $v_{\text{data}} = [|\psi_i|]_{i=0}^{15}$. Here, $\tilde{\psi}_2, \tilde{\psi}_3$ denotes modified values of $\psi_2, \psi_3$ in the data vector. Shift operations are given by $O_{\text{shift}}(k, d)$ (refer \cref{eq:O_shift}). Deletion operations are given by $O_{\text{del}}(k, r) = D^{|k\rangle}_{r_i}$ (refer \cref{eq:O_del}). Similarly, insert operations are represented as $I^{|k\rangle}_{r_{i}}$.}
	\label{table:Matrix}
\end{table}

\begin{table}[ht]
	\centering
	\begin{tabular}{| l | l |}
		\hline
		\multicolumn{2}{|c|}{Common shift operators}\\
		\hline
		$L(k, 0)$ &$(\psi_0, 0000), (\psi_1, 0001), (\psi_7, 0111), $\\
		&$(\psi_8, 1000), (\psi_{11}, 1011), (\psi_{12}, 1100),$\\
		&$(\psi_{14}, 1110), (\psi_{15}, 1111)$\\
		\hline
		$L(k, 1)$ &$(\psi_6, 0110), (\psi_8, 1000), (\psi_9, 1001),$\\
		&$(\psi_{11}, 1011)$\\
		\hline
		$L(k, 2)$ & $(\psi_0, 0000), (\psi_6, 0110), (\psi_9, 1001),$\\
		&$(\psi_{12}, 1100)$\\
		\hline
		$L(k, 3)$ & $(\psi_7, 0111), (\psi_8, 1000), (\psi_9, 1001),$\\
		&$(\psi_{13}, 1101)$\\
		\hline
		$L(k, 4)$ & $(\psi_{10}, 1010), (\psi_{11}, 1011), (\psi_{12}, 1100),$\\
		&$(\psi_{13}, 1101)$\\
		\hline
		$R(k, 0)$ & $(\psi_4, 0100), (\psi_5, 0101)$\\
		\hline
		$R(k, 2)$ & $(\psi_5, 0101)$\\
		\hline
	\end{tabular}
	\caption{Common shift operations of the data elements \cref{table:Matrix}. Here $(\psi_k, |k\rangle)$ denotes $k^{\text{th}}$ data element in $|k\rangle$ basis state represented in binary. Note that the $L(k, 0)$ shift involves six data elements; however, as illustrated in \cref{example_3}, the zeros in $\psi_{14}$ and $\psi_{15}$ can be exploited to extend this to eight shifts.}
	\label{table:Common_shift}
\end{table}

\subsection{Structured Real Matrix}\label{sec:Sparse_real_matrix}
\begin{figure*}[t]
	\centering
	\begin{subfigure}[t]{0.25\textwidth}
		\centering
		\scalebox{0.8}{
			\def\svgwidth{\textwidth}
			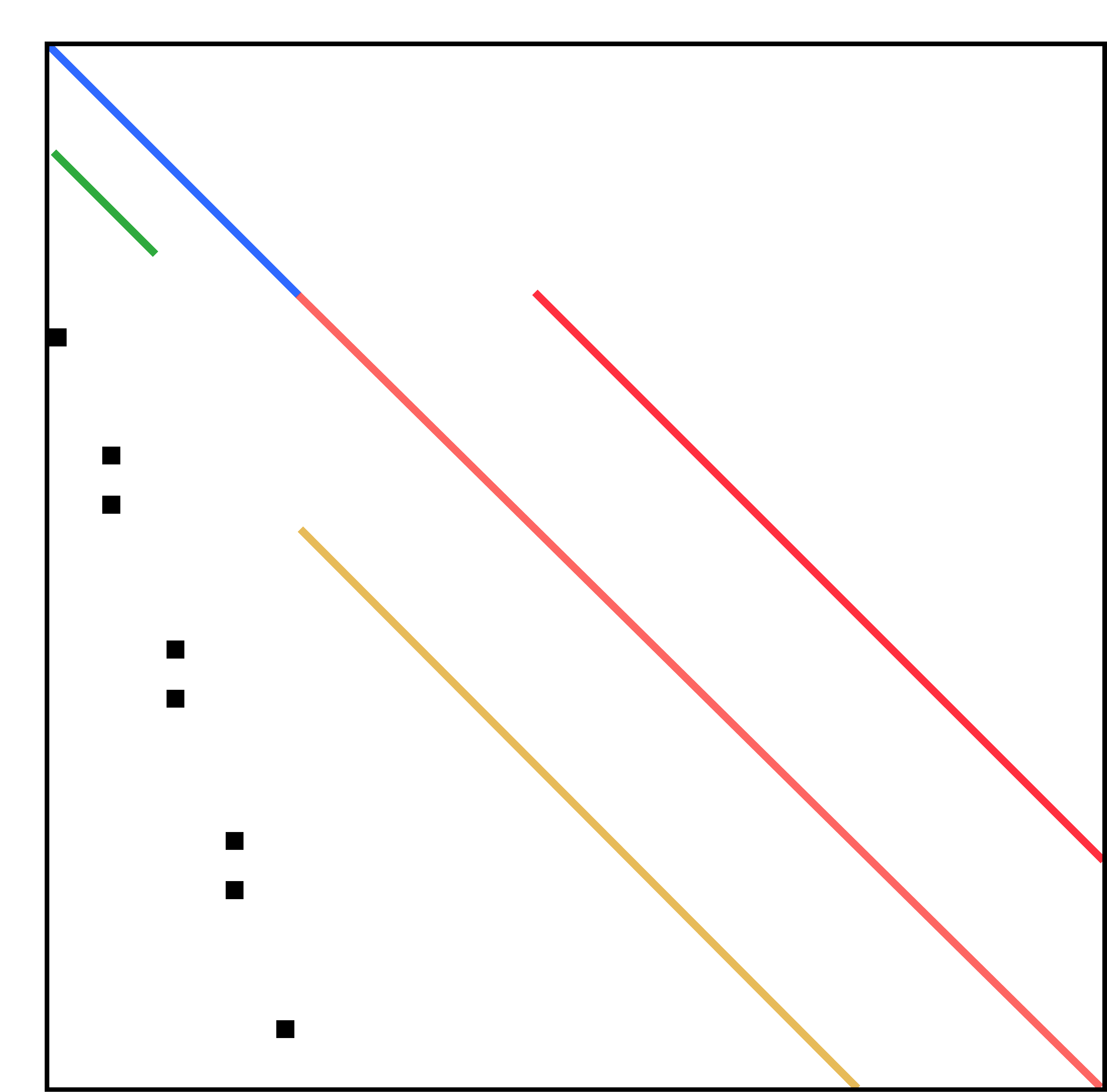
		}
		\caption{}
		\label{fig:8a_Matrix}
	\end{subfigure}
	\begin{subfigure}[t]{0.37\textwidth}
		%\centering
		\includegraphics[width=\textwidth]{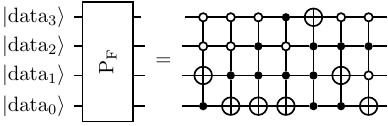}
		\caption{}
		\label{fig:8b_Matrix}
	\end{subfigure}
	\begin{subfigure}[t]{0.33\textwidth}
		%\centering
		\includegraphics[width=\textwidth]{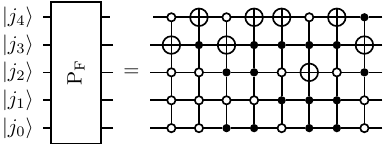}
		\caption{}
		\label{fig:8c_Matrix}
	\end{subfigure}
	\label{fig:8_Matrix}
	\caption{\subref{fig:8a_Matrix} Visual representation of the example sparse matrix structure (see \cref{sec:Sparse_real_matrix}). The data elements are denoted by $[\psi_i]_{i=0}^{13}$. Continuous lines of the same color indicate the repeating pattern of a data element $(\psi_i)$ along a diagonal $d$, while black square dots denote single, non-repeating data elements. \subref{fig:8b_Matrix} Coherent permutation on the data qubits \cref{eq:right_permute_mapping}. \subref{fig:8c_Matrix} Coherent permutation on the matrix qubits \cref{eq:left_permute_mapping}.}
\end{figure*}

Consider a sparse real matrix $A = \mathbb{R}^{32 \times 32}$ with the structure shown in \cref{fig:8a_Matrix}. Following the block-encoding procedure outlined in \cref{sec:Complete_circuit}, the corresponding data vector is $v_{\text{data}}=[|\psi_i|]^T,\; i\in[0,13]$ and sign vector is $v_{\text{sign}}=[\operatorname{sgn}(\psi_i)]^T,\; i\in[0,13]$.
Since $\dim(v_{\text{data}})=14$, we require $m=\lceil\log_2(14)\rceil=4$ data qubits, and the remaining two basis states are introduced as zero-amplitude padding states.

The block-encoding operations (shift, delete, insert) are summarized in \cref{table:Matrix} (refer \cref{sec:Optimized_Index_Mapping_Oracle}). Note that two distinct values $(\psi_2, \psi_3)$, occur on the main diagonal. Within the block-encoding framework, these appear as the combined diagonal value $\psi_2 + \psi_3$ (see \cref{eq:diagonal_block_encoded}). To address this case, we outline three possible encoding strategies with an objective to reduce MCX gates and subnormalization factor:
\begin{enumerate}
	\item Without modification: $\psi_2, \psi_3$, \\
	Delete operations: $\psi_2: D^{|0010\rangle}_{\{5-31\}}, \psi_3: D^{|0011\rangle}_{\{0-4\}}$,\\ 
	Contribution to $\alpha$: $|\psi_2| + |\psi_3|$.
	\item With modification: $\tilde{\psi}_2 = \psi_3 - \psi_2, \tilde{\psi}_3 = \psi_2$,\\
	Delete operations: $\tilde{\psi}_2: D^{|0010\rangle}_{\{0-4\}}, \tilde{\psi}_3: \text{Nothing}$,\\ Contribution to $\alpha$: $|\tilde{\psi}_2| + |\tilde{\psi}_3|$, \\
	This is advantageous when $|\psi_3 - \psi_2| < |\psi_3|$.
	\item With modification: $\tilde{\psi}_2 = \psi_3, \tilde{\psi}_3 = \psi_2 - \psi_3$, \\
	Delete operations: $\tilde{\psi}_2: \text{Nothing}, \tilde{\psi}_3: D^{|0011\rangle}_{\{5-31\}}$,\\
	Contribution to $\alpha$: $|\tilde{\psi}_2| + |\tilde{\psi}_3|$,\\
	This is advantageous when $|\psi_2 - \psi_3| < |\psi_2|$.
\end{enumerate}
For demonstration, we choose the second approach as illustrated in \cref{table:Matrix}. Next, we determine the common shift operators (refer \cref{sec:Optimized_Index_Mapping_Oracle}), shown in \cref{table:Common_shift}.

\graphicspath{{./Images/8/}}

We demonstrate the optimized index mapping oracle for the block encoding operations in \cref{table:Matrix}. For demonstration purposes, we consider the shift $L(k, 0)$ in \cref{table:Common_shift}. For $L(k, 0)$, the control states $S_2$ does not satisfy the structural requirements in \cref{theorem:MCX_Composition}. Therefore, we determine $F=\{|\text{data}_0\rangle\}$ and apply the permutation operator $\text{A\_PERMUTE}_\text{F}$ as shown in \cref{fig:5a_Reduction}. Following \cref{theorem:Mapping}, the mapping $\phi: S_2 \mapsto S_3$ is obtained by solving the linear sum assignment problem \cite{crouse2016implementing}. Finally, the gates are generated to permute $(S_2 \mapsto S_3)$ according to \cref{definition:Permutation}, resulting in the following local walk operators:
\begin{equation}\label{eq:right_permute_mapping}
\begin{aligned}
	\psi_0\hspace{0.2cm}&0000\\
	\psi_1\hspace{0.2cm}&0001&\rightarrow0011&\rightarrow0010\\
	\psi_7\hspace{0.2cm}&0111&\rightarrow0110\\
	\psi_8\hspace{0.2cm}&1000\\
	\psi_{11}\hspace{0.2cm}&1011&\rightarrow1010\\
	\psi_{12}\hspace{0.2cm}&1100\\
	\psi_{14}\hspace{0.2cm}&1110\\
	\psi_{15}\hspace{0.2cm}&1111&\rightarrow0111&\rightarrow0101&\rightarrow0100
\end{aligned}.
\end{equation}
The MCX gates implementing the mapping in \cref{eq:right_permute_mapping} (see \cref{theorem:A_PERMUTE}) correspond to the operator $\text{A\_PERMUTE}_{\text{F}}$, as illustrated in \cref{fig:8b_Matrix}. Alternatively, one may employ multi-swapping strategies as discussed in \cref{theorem:MCX_Inverse}. 

Considering the combined delete $D^{|0001\rangle}_{\{0,5,10,15,20,25,30,31\}}$ (refer \cref{table:Matrix}), the control states $S_2$ require permutation. Therefore, we determine $F = \{|j_4 j_3\rangle\}$ and implement $\text{A\_PERMUTE}_\text{F}$, as shown in \cref{fig:5d_Reduction}. Following the same protocol as for the previous case, we obtain the mapping $\phi: S_2 \mapsto S_3$, leading to the following local walk operators:
\begin{equation}\label{eq:left_permute_mapping}
	\begin{aligned}
		0\hspace{0.2cm}&00000&\rightarrow01000&\rightarrow11000\\
		5\hspace{0.2cm}&00101&\rightarrow01101&\rightarrow11101\\
		10\hspace{0.2cm}&01010&\rightarrow11010\\
		15\hspace{0.2cm}&01111&\rightarrow01011&\rightarrow11011\\
		20\hspace{0.2cm}&10100&\rightarrow11100\\
		25\hspace{0.2cm}&11001\\
		30\hspace{0.2cm}&11110\\
		31\hspace{0.2cm}&11111
	\end{aligned}.
\end{equation}
The MCX gates corresponding to the mapping in \cref{eq:left_permute_mapping} (see \cref{theorem:A_PERMUTE}) are implemented through $\text{A\_PERMUTE}_{\text{F}}$, as illustrated in \cref{fig:8c_Matrix}. The complete circuit for block encoding the matrix \cref{fig:8a_Matrix} is obtained by combining the shift, delete, insert, and permutation operators, as shown in \cref{fig:6a_Complete_circuit}. This circuit provides a practical gate-level realization that can be directly employed within quantum algorithms. 

\section{Discussion}

In this work, we developed a framework for block encoding of sparse matrices with hardware-aware circuit constructions and local reductions in MCX control complexity. The construction provides an explicit bridge from oracle-level block encoding to gate-level implementations. Using existing state-preparation primitives, we focus primarily on optimizing index-mapping oracles, including shift, delete, insert, and coherent permutation operations.

The framework allows an entry-wise construction for general sparse matrices. However, its principal efficiency benefits are expected for exploitable structure. Highly irregular sparse matrices with predominantly distinct entries and non-repetitive diagonal structure may exhibit higher index-mapping overhead due to limited opportunities for MCX compression.

At the circuit level, we show that several shift, delete, and insert operations can be combined through MCX compression when the associated control structures admit favorable grouping. When such direct compression is not available, we formulate the resulting basis-state reordering as a coherent permutation problem and propose a routing-based construction. This yields an optimized index-mapping oracle that is compatible with hardware-aware circuit synthesis and can reduce long-range interactions by enabling nearest-neighbor–friendly MCX structures under appropriate connectivity assumptions. However, the overall resource cost depends not only on MCX compression but also on the decomposition of MCX gates into native operations and the routing overhead induced by the underlying hardware connectivity graph.

The hardware resource analyses on representative examples (see \cref{table:resource_analysis_example_2,table:resource_analysis_example_4,table:resource_analysis_example_5}) indicate that the proposed optimization framework can lead to significant reductions in circuit depth. The complete block-encoding construction was illustrated through two circuit architectures, and representative examples were used to demonstrate the end-to-end pipeline from oracle-level formulations to explicit gate-level implementations. The resulting circuits are directly applicable to key quantum algorithms—including generalized frameworks such as QSVT as well as specific applications like Hamiltonian simulation and quantum linear system solvers. 

\section*{Acknowledgements}
This research was funded through the European Union’s Horizon Programme (HORIZONCL4-2021-DIGITALEMERGING-02-10), Grant Agreement 101080085 (QCFD).

\bibliography{bibliography.bib}

\end{document}